\documentclass[aps,prd,twocolumn,nofootinbib,superscriptaddress,showpacs,amsmath,amssymb]{revtex4}
\usepackage{graphicx,epsf}
\usepackage[]{latexsym}

\newcommand{\be}{\begin{eqnarray}}
\newcommand{\ee}{\end{eqnarray}}
\newcommand{\bea}{\begin{eqnarray}}
\newcommand{\eea}{\end{eqnarray}}

\def\comment#1{}

\usepackage{color}

\definecolor{darkred}{rgb}{.8,0,0}

\definecolor{darkblue}{rgb}{0,0,.7}

\definecolor{darkgreen}{rgb}{0,.7,0}

\begin{document}

%
%
\title{Exponential distance relation (aka Titius-Bode rule) in extra solar planetary systems}

\author{Dimitrios Krommydas}
\email{krommydas.di@gmail.com}
\affiliation{Institute-Lorentz for Theoretical Physics, Leiden University, \\P.O. Box 9506, Leiden, The Netherlands}
\author{Fabio Scardigli\footnote{corresponding author}}
\email{fabio@phys.ntu.edu.tw}
\affiliation{Institute-Lorentz for Theoretical Physics, Leiden University, \\P.O. Box 9506, Leiden, The Netherlands}
\affiliation{Dipartimento di Matematica, Politecnico di Milano, Piazza Leonardo da Vinci 32, 20133 Milano, Italy}


%
%
%
%
%
%
%
%
%
\begin{abstract}
\par\noindent
 In this paper we present phenomenological evidence for the validity of an exponential distance relation (also known as generalized Titius-Bode law) in the 32 planetary systems (31 extra solar, plus our Solar System) containing at least 5 planets each (known up to July 2023). We produce the semi-log fittings of the data, and we check them against the statistical indicators of $R^2$ and $Median$. Then we compare them with the data of 4000 artificial planetary systems created at random. In this way, a possible origin by chance of the Titius-Bode rule (TBR) is reasonably excluded. We also point out that in some systems the fittings can be definitely improved by the insertion of new planets into specific positions. We discuss the Harmonic Resonances method and fittings, and compare them with the Titius-Bode fittings. Moreover, for some specific systems, we compare the Titius-Bode fitting against a polynomial fitting ($r\sim n^2$). Further comparisons with previous relevant works are reported in the last section. It emerges that TBR describes 25 out of the 32 planetary systems ($78\%$) with a $R^2\geq 0.95$. Further, it results to be the most economical (in terms of free parameters) and best fitting law for the description of spacing among planetary orbits. This analysis allows us to conclude that an exponential distance relation can reasonably be considered as ``valid'', or strongly corroborated, also in extra solar planetary systems. 
\end{abstract}
\pacs{N/A. Key words: Planetary Systems, exoplanets}

\maketitle

\section{Introduction}


Following the footsteps of the visionary work of Johannes Kepler (Kepler 1596), for more than 170 years scholars had been looking for a law able to encode in a formula the distances of the known planets from the Sun. Kepler proposed the platonic solids as a guide to the numerical progression of the major axis of the elliptic orbits he envisaged. Such an idea was quite rapidly abandoned, especially after the dynamics of the Solar System had been unveiled by Isaac Newton. 
Nevertheless the quest for a distance relation among the planets of the Solar System remained alive. And actually the quest soon extended also to the systems of moons of Jupiter and Saturn, discovered in the meanwhile. After several precursors had paved the way (among them we count Christian von Wolff, mathematician, physicist, philosopher, and his brilliant disciple Immanuel Kant), finally a rule for the distances of planets from the Sun was proposed by Johann Daniel Titius (Titius 1766), and published as a note in his German translation of the Charles Bonnet's \emph{Contemplation de la Nature} (1764). Soon the law was noticed and popularized by the much more famous astronomer Johann Elert Bode (Bode 1772). 

The original formulation of the rule, proposed by Titius, can be expressed by the simple formula
\be
r(n) = 0.4 + 0.3\cdot 2^n \,.
\label{TBold}
\ee
where the distance $r(n)$ of the planet from the Sun is given in astronomical units, i.e. in terms of the 
radius of the Earth's orbit (which defines 1 Astronomical Unit $\simeq 150 \times 10^6$ km). For $n=-\infty, 0, 1, 2,...$ this relation gives the distances $r(n)$ from the Sun, respectively, of Mercury, Venus, Earth, Mars, etc., including the asteroid belt (actually, Ceres, the heaviest asteroid, was discovered by Piazzi in 1801 by following the indication of this rule with $n=3$), and Uranus ($n=6$), which at the moment of the first formulation of the law (1766-1772) had not yet been discovered 
(see the book of Nieto (1972) for history, different explicit formulations, and extensive bibliography).

In its original form (\ref{TBold}), the relation was not able
to account for the distance of Neptune and Pluto. During the 20th century more refined
versions of the rule were elaborated (Blagg 1913; Richardson 1943; Dermott 1968). The present-day versions are able to describe not only the planetary distances within the solar system, including planets like Neptune
and Pluto, but also can be successfully applied to the systems of
satellites orbiting Jupiter, Saturn and Uranus (See Appendix D for the precise ways 
in which those moons obey TBR). The agreement
between the predicted and the observed distances of the various
satellites from the central body is really good, of the
order of a few percents (see e.g. again Nieto 1972). 
The modern version of the Titius-Bode rule can be expressed, if we neglect
second order corrections, by an exponential relation as
\be
r(n) = a\,e^{2\lambda n},
\label{tbl}
\ee
where the factor $2$ is introduced for convenience reasons and $n=1,\, 2,\, 3,\, \dots$.\\
For the Solar System we have
\be
2 \lambda &=& 0.53707, \quad \quad \quad  e^{2\lambda} \simeq 1.7110, \nonumber \\
a &=& 0.21363 \,\,\, {\rm A.U.} \nonumber
\ee
The amazing thing found by Blagg was that
the geometric progression ratio $e^{2\lambda}$ is roughly the same
for the Solar System and for the satellite
systems of Jupiter ($e^{2\lambda} \simeq 1.7277$), Saturn ($e^{2\lambda} \simeq 1.5967$),
and Uranus ($e^{2\lambda} \simeq 1.4662$). The parameter $\lambda$ is dimensionless,
its value is inferred from the observed data, and it depends on the specific system considered
(planetary or satellite system).
Also the parameter $a$ is in general obtained from observations, it has the dimension of a length and it is linked to the radius of the first orbit of the system considered, since $r(1)=ae^{2\lambda}$. With a slightly different formulation of the TB rule as $r(n)=a e^{2\lambda(n-1)}$, $n=1,2,3,\dots$, then $a$ coincides with the radius of the first orbit, $r(1)=a$.\\

Despite its quite evident successes in describing the spacing of the local planetary systems (Solar system, plus the satellites systems of Jupiter, Saturn, Uranus), the physical interpretation of the TB relation has been often questioned and remains a matter for heated debate. 

Broadly speaking, there are at least three groups of opinions regarding the TBR and its physical significance: 
(i) the TBR constitutes just simply sheer numerology; 
(ii) the TBR is valid only for some specific systems due to particular conditions occurred during the planet formation process;
(iii) the TBR is largely a direct consequence of (some) planetary stability requirements which should be satisfied during the course of the systems’ existence.

Many criticisms have been risen against an effective physical meaning of the law. 
For example, according to Graner \& Dubrulle (1994) the TB relation is probably just a consequence of the scale invariance of the disk which gave rise to the planets. However, according to Lynch (2003), it is not possible to conclude unequivocally that laws of Titius–Bode type are, or are not, physically significant. In other words, Lynch convincingly argued that the agreement with the observations cannot be safely considered as a mere statistical chance. So, the possibility of a physical explanation for the observed distributions remains open. 

A somehow similar conclusion also appears to be consistent with the work by Hayes and Tremaine (1998). Following an approach which involves some statistical numerical experiments, they chose to fit randomly selected artificial planetary systems to Titius-Bode type laws by considering a distance rule inspired by the Hill stability of adjacent planets. They did not identify a particularly profound significance of the TBR, except that its general meaning is that stable planetary systems tend to be spaced in a regular manner.

On the side of physical models, many theories have been developed during the last 250 years to explain the Titius-Bode relation.
There have been dynamical models connected with the theory of the
origin of the solar system (Alfven 1954), 
gravitational theories (Schmidt 1946), nebular theories (Weizs\"acker 1943, Willerding 1992), just to cite some of them. 
Many of them have been excellently reviewed in the book of Nieto (1972).

Also the approaches involving ideas from scale relativity, or stochastic trajectories, 
or also Schr\"odinger-like equations, in order to give
account of the rule (\ref{tbl}) have a robust tradition. During the years many authors
have suggested various models in this direction (for an incomplete list of
papers see for example: Caswell 1929; Albeverio et al. 1983; Nottale 1996; Nottale et al. 1997; Agnese \& Festa 1997; Reinisch 1998; de Oliveira Neto et al. 2004; Scardigli 2007).


Of course, in the last 15-20 years, with the discovery of a growing number of extra solar planetary systems, it has become increasingly important to check the TB relation in the new observed systems.
Following the original tradition, many authors have applied the TB rule to exoplanetary systems in order to predict new planets. In this direction goes for example the paper of Bovaird \& Lineweaver (2013). 
The exoplanetary system 55 Cancri (HD 75732), containing five planets, was investigated in some detail using the TBR by Chang (2008), Poveda \& Lara (2008), and Cuntz (2012). 
Cuntz (2012) in particular argued that new planetary candidates were predicted in the 55 Cancri system 
through the Titius-Bode’s relation, and perhaps one of the new planets could be habitable. Also the planetary system revolving around the star HD10180 has been studied in some specific detail (Lovis et al. 2011), and it appears to satisfy the TB rule.
According to Chang (2010) one cannot rule out the possibility that the distribution of the ratio of orbiting periods in multiple planet systems is consistent with that derived from Titius-Bode’s relation.
Lara et al. (2020) used data from 27 exoplanetary systems
with at least 5 planets and showed that the planetary orbital
periods in exoplanetary systems are not consistent with a
random distribution.\\
%
%
%
  
The main purpose of the present paper is to examine the 32 planetary systems that, up to today (July 2023, see NASA Archive), appear to host at least 5 planets each, all orbiting around a \textit{single} star, and to see if and to what extent the TB rule is satisfied in such systems. We decided to consider systems with at least 5 planets in order to fit at best the two parameters of TB rule, and to minimize the possible (statistical) errors. Of course, this investigation has been made possible by the surveys and discoveries operated by Kepler and TESS satellites, in particular in the last ten years or so.\\
On the other hand, a survey aimed at checking the TB rule in extra solar planetary systems seems to be quite timely now, since the data released by Kepler and TESS satellites cover a significant number of systems, each one endowed with a significant number of confirmed planets (at least five, as said). \\
As for the source of our data, we decided to stick with the NASA Exoplanet Archive (NASA Archive), in order to have an up-to-dated source, as well as to enjoy a reliable uniformity in the presentation of the data.\\   
In this paper we focus mainly on an exponential spacing law (aka \textit{Titius-Bode rule}). In Sec.II we present the planetary systems under scrutiny, and discuss the analytical tools employed to produce fittings and graphs. In Sec.III (following the tradition which led to the proposal and discovery of the asteroid belt), we present the possibility to improve some fittings by inserting new planets to fill suspected gaps among existing planetary orbits. We discuss several examples and propose some ``predictions". In Sec.IV we discuss a possible relation between the TB rule and the age of the planetary systems. In Sections V and VI we briefly discuss also different descriptions of the planetary spacing data, namely the harmonic resonances fitting (Sec.V), and the polynomial fitting (Sec.VI). Finally, Sec.VII is devoted to some further specific comparisons with previous works, and then to Conclusions.\\      
We are pretty confident in saying that the main result of our paper is that the TB law seems to be confirmed as the 
``best" rule to describe planetary spacing also among extra solar planetary systems, at least for what concerns the best fitting obtained with a minimum number of free parameters (two).



%
%
%

\section{Fittings and Graphs}
%


The list of planetary systems considered in this work is given in Table \ref{table1}. 

\begin{table}[!ht]
\centering
\begin{tabular}{|c|c|c|c|c|}
	\hline
	System & Planets & $ R^2 $ & Median & Notes\\
	\hline
    Sun & 10 & 0.993448869 & 0.019581007 & with Ceres\\
	\hline
	KOI-351 &  8 & 0.96322542 & 0.059404861 & Kepler 90\\
	\hline
	Trappist 1 & 7  & 0.994255648 & 0.012727939 &\\
	\hline
	HD 10180 &  6  &  0.992725066 &	0.06644463 &\\
	\hline
	HD 191939 &  6  &  0.945551901 &	0.08825776 & 2022\\
	\hline
	HD 219134 &  6  &  0.913657699 &	0.126773358 &\\
	\hline
	HD 34445 & 6  &  0.974290137	& 0.108819429 &\\
	\hline
	K2-138 &  6  &  0.945458309 & 0.094321758 & 2021\\
	\hline
	Kepler 11 &  6 & 0.962373014 &	0.044210656 &\\
	\hline
	Kepler 80 &  6 & 0.955414548	& 0.04023616 &\\
	\hline
	TOI - 1136 &  6 & 0.987422135 & 0.031168586 & 2022\\
	\hline
	TOI - 178 & 6 & 0.983217838  & 0.047680406  & 2021\\
	\hline
	HD 108236 &  5 & 0.976381637	& 0.037194964 & 2021\\
	\hline
	HD 158259 &  5  & 0.999713447	& 0.003377513  & 2020\\
	\hline
	HD 23472 &  5  & 0.988496548 & 0.022330332 & 2022\\
	\hline
	HD 40307 &  5 &  0.984933377 & 0.046130252 &\\
	\hline
	K2-268 & 5 &  0.957182918 &  0.062647955  & 2019\\
	\hline
	K2-384 & 5 & 0.983020508 & 0.04991254 & 2022\\
	\hline
	Kepler 102 & 5  &  0.987900552 &	0.017713236 &\\
	\hline
	Kepler 122 & 5 &  0.98710756 &  0.026533536 &\\
	\hline
	Kepler-150 & 5  & 0.8605413 &	0.236636167 &\\
	\hline
	Kepler 154 & 5 & 0.985703723 & 0.045352455 &\\
	\hline
	Kepler 169 &  5 &  0.875354644 &	0.151151179&\\
	\hline
	Kepler 186 &  5 &  0.92681234 &	0.093772029&\\
	\hline
	Kepler 238 & 5 & 0.98771232 & 0.053557294 &\\
	\hline
	Kepler 292 & 5 &  0.995427692 &	0.008569513 &\\
	\hline
	Kepler 32 &  5 &  0.971262925 &	0.042153715 &\\
	\hline
	Kepler 33 &  5 &  0.948841017 & 0.054322078 &\\
	\hline
  Kepler 55 & 5 & 0.988367826 &  0.02874602 &\\
	\hline
Kepler 62 & 5 &  0.950734271 & 0.15977283 &\\
	\hline
Kepler 82 & 5 & 0.955569351  & 0.086643589  & 2019\\
	\hline
Kepler 84 & 5 &  0.992394962	& 0.022666502 &\\
	\hline
\end{tabular}
\caption{Each of the 32 systems considered in the present work hosts at least 5 planets, revolving around a single star. The statistical indicators displayed are $R^2$ and $Median$ (for definitions see main text and appendices). We have a maximum agreement when $R^2_{max}=1$ and $Median_{max}= 0$. For references on each single planetary system see Table \ref{table14}.}
\label{table1}
\end{table}
As we said, the data used to construct Table \ref{table1} are taken from NASA Exoplanet Archive\footnote{For this reason we excluded the system HIP 41378, which seems to have 5 planets, but the data of the fifth planet are not available on the NASA Archive.}. Hereafter we discuss the reasons to consider that list, give comments and explanations of parameters appearing in Table \ref{table1}, and discuss anomalous cases.

Firstly, in our analysis, we only consider extra solar planetary systems with at least \textit{five} confirmed planets each (or more), revolving around a \textit{single} star. The choice of the number ``5" for the population of the planetary systems examined, could appear quite arbitrary, and perhaps it is. However, simple practical considerations push towards that choice. Since our aim is to fit the distances/periods of the planets of a single system with a two-parameters exponential law, of course one would like to maximize the number of points contained in a single system, and therefore to choose systems with a large number of planets. To compute with accuracy the coefficients of a semi-log linear regression requires as many points as possible. That is even more true, if we consider that some of the TB regression lines will be used to ``predict" the presence of new planets in a given system. Obviously, the smaller the number of points you have at your disposal, the less performing is an exponential TB law in this prediction task.  
On the other hand, the known systems with many planets are not so many, and we also want to check the TB relation in as many different systems as possible, in order to strongly corroborate its statistical validity. The balance between these two opposite requirements has pushed us to choose systems populated with at least 5 planets.
   
For sake of completeness, in Table \ref{table2} we report also the statistical analysis for planetary systems containing at least 5 planets, but with 2 or 3 stars at their center. Although some of these systems present very good statistical indicators (with the exception of GJ 667 C), it is also fair to say that the presence of 2 or 3 stars at the center of these systems can complicate the understanding of their dynamics in unexpected ways. Therefore, as for the discussion of the phenomenology of the Titius-Bode rule in extra solar planetary systems, we think it is prudent to confine ourselves to single-star systems, at least in this paper, postponing the analysis of multiple-star systems to future works \footnote{Further, we note that the system HD 20781-2, endowed with 5 planets, has not been reported neither in Table \ref{table1} nor in Table \ref{table2}, since not only it contains 2 stars at its center, but apparently \textit{four} planets revolve around HD 20781, and \textit{one} planet revolves around HD 20782... a really too complicated system!}.

\begin{table}[!ht]
\centering
\begin{tabular}{|c|c|c|c|c|}
	\hline
	System & Planets & $ R^2 $ & Median & Notes\\
	\hline
  Kepler 20 & 6 &  0.994934868 &	0.020480257 & 2 stars\\
	\hline
	55 Cancri & 5  & 0.974846945 & 0.105266615 & 2 stars\\
	\hline
	GJ 667 C & 5 & 0.940319108 & 0.085544299 & 3 stars\\
	\hline
	Kepler 296 & 5 &  0.999532566 &	0.006855603 & 2 stars\\
	\hline
  Kepler 444 & 5 &  0.996885042 &	0.007292345 & 3 stars\\
	\hline 
\end{tabular}
\caption{$R^2$ and $Median$ analysis for planetary systems with 2 or 3 stars at their center. Maximum agreement for $R^2_{max}=1$ and $Median_{max}= 0$.}
\label{table2}
\end{table}




\subsection{Semi-log formulation of TB rule}


In Table \ref{table3} we display a sample of the raw data of one of the 32 planetary systems under our scrutiny (precisely, KOI-351), data that we used to build the semi-log linear regressions, and hence to compute the $R^2$ and $Median$ statistical indicators showed in Table \ref{table1}. In particular, to our scope, pivotal data are: number of planets for each system; number of stars in each system (we considered planetary systems with only one single star at the center); period and/or semi-major axis of each planetary orbit.

\begin{table}[!ht]
\centering
\begin{tabular}{|c|c|c|c|}
	\hline
  Planet &  Host  & N.Stars \&   & Orbital Period \\
	 name  &  name  & N.Planets    & (days)\\
	\hline
	KOI-351 b &  KOI-351  & 1 - 8 & 7.008151$\pm$0.000019 \\
	\hline
	KOI-351 c &  KOI-351  & 1 - 8  & 8.719375$\pm$0.000027 \\
	\hline
	KOI-351 d &  KOI-351  & 1 - 8  & 59.73667$\pm$0.00038 \\
	\hline
	KOI-351 e &  KOI-351  & 1 - 8  & 91.93913$\pm$0.00073 \\
	\hline
	KOI-351 f &  KOI-351  & 1 - 8  & 124.9144$\pm$0.0019 \\
	\hline
	KOI-351 g &  KOI-351  & 1 - 8  & 210.60697$\pm$0.00043 \\
	\hline
	KOI-351 h &  KOI-351  & 1 - 8  & 331.60059$\pm$0.00037 \\
	\hline
	Kepler-90 i & KOI-351 & 1 - 8  & 14.44912$\pm$0.00020 \\
	\hline
	\end{tabular}
\caption{Raw data for the planetary system KOI-351.}
\label{table3}
\end{table}

In order to produce a linear fitting of the data presented in Table \ref{table3}, it is useful to take the 
$log$ of relation (\ref{tbl}) so that
\be
\log r(n) = \log a + 2\lambda n
\label{logTB}
\ee
which can be rewritten as
\be
Y(n) = A + B n 
\ee
where $Y(n)=\log r(n)$. The linear fittings will provide the coefficients $A$ and $B$, which are linked to the TB parameters by the relations
\be
a&=&e^A \nonumber \\
\lambda &=& B/2\,.
\ee
Since the planetary orbital periods are actually the variables directly measured by the observers, instead of semi-major axes, we can make use of Kepler Third Law, $T^2 =k r^3$, to reformulate the rule (\ref{tbl}) in terms of orbital periods and integers
\footnote{A formulation of the exponential TB relation that uses orbital periods (in place of semi-major axis) is also known under the name of ``Dermott's law" (Dermott 1968).}
\be
T^2 = k r^3 = k a^3 e^{6\lambda n}
\ee
that is
\be
T(n) = (k a^3)^{1/2} \, e^{3\lambda n}\,.
\ee
Taking the $log$ of both members we have 
\be
Z(n) = C + Dn
\label{logTBT}
\ee
where 
\be
Z(n)=\log T(n)\,, \quad C = \frac{1}{2}\log (ka^3)\,, \quad D=3\lambda\,.
\label{Z}
\ee

\begin{figure}
\includegraphics[width=\linewidth]{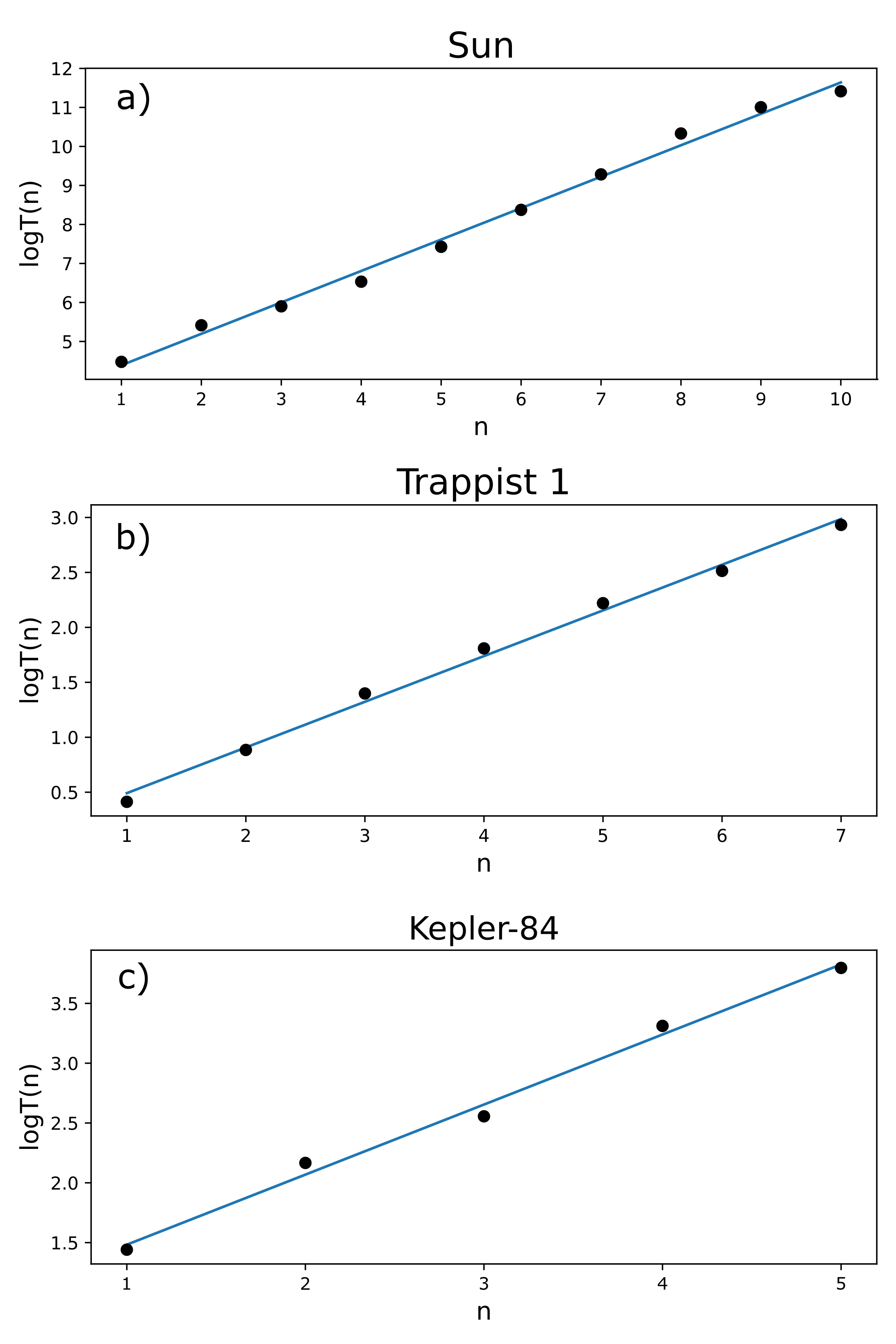}
\caption{Examples of various TB fittings for different systems.\\
\textit{a}) TB fitting of the Solar system, including Ceres and Pluto, namely 10 planets. The statistical indicators (see Table~\ref{table1}) are $R^2=0.993448869$ and $Median = 0.019581007$.\\
\textit{b}) TB fitting of Trappist 1, a system with 7 confirmed planets. The statistical indicators (Table~\ref{table1}) are $R^2=0.994255648$ and $Median = 0.012727939$.\\
\textit{c}) TB fitting of Kepler-84, a system with 5 confirmed planets. The statistical indicators (Table~\ref{table1}) are $R^2=0.992394962$ and $Median = 0.022666502$.}
\label{Fig3_(18-6-23)_V2.png}
\end{figure}

Following Eqs.\eqref{logTB}-\eqref{logTBT}, for each planetary system we proceed to order in an increasing way the periods $T(n)$, or better the $\log T(n)$, and we label each term of the resulting sequence with an increasing natural integer 
$i=1, 2, 3, \dots$. In so doing we obtain, for each system, the set of data $\{i, \log T(i) \}$ to which the usual linear regression method is applied (the \textit{least squared method}). Some of these plots are displayed, as examples, at the end of this Section in Fig.\ref{Fig3_(18-6-23)_V2.png}. 



In our diagrams we do not report error bars. The reason is simple: from Table \ref{table3}, and from the general tables in 
the NASA Archive, it appears clearly that the large majority of the errors on the periods (measured in days) are at most of order $10^{-2}$ or less. Since we are interested in displaying on the diagrams the logs of periods, any (already small) error bar would be further strongly suppressed. In formulae, if we have $T=T_0 \pm \varepsilon$, then
\be
\log(T) = \log (T_0 \pm \varepsilon) \simeq \log T_0 \pm \frac{\varepsilon}{T_0}\,,
\ee   
and displaying error bars of order $10^{-3}$ or less would be graphically tough, as well as useless.

\subsection{Statistical tools}

The tools we use to quantify the ``goodness" of the linear regressions, and consequently of the evaluations of the TB parameters, are the $R^2$ and $Median$ statistical indicators (see Appendices A, B, C for definitions and properties).
An inspection of Table \ref{table1} reveals that, of the 32 single-star planetary systems (including our Solar systems), at least 6 have an $R^2$ greater than 0.99, revealing a very good agreement between the astronomical data and the linear regression with the phenomenological TB rule.

One of the most common arguments used against the TB rule concerns a possible origin at random of the law itself. According to this point of view, it would be statistically "easy" to produce planetary systems at random, and they would naturally turn out to obey the TB relation, by pure chance.
To explore this possibility, we created randomly 4000 artificial planetary systems: 1000 systems with 8 planets each, 1000 with 7, 1000 with 6, and 1000 systems with 5 planets each, respectively. In this first naive attempt we did not impose any particular further constraint. Essentially, referring for example to a system with 8 planets, 
we extracted at random 8 numbers $T_i$, with $0 \leq T_i \leq 1$, namely the periods of the 8 planets, and reordered them in an increasing way, so that $T_i<T_j$ iff $i<j$. Of course, through a trivial rescaling $\mu T_i$, these numbers can represent any period from zero to infinity. We then fit the points $\{i, \log T_i\}$, $i=1,2,\dots,8$, with a linear regression, and compute the relevant $R^2$ and $Median$ indicators. The averages over the 1000 random systems are displayed in Table \ref{table4}. 

\begin{table}[!ht]
\centering
\begin{tabular}{|c|c|c|c|}
	\hline
	Systems with& $ R^2 $ & Median & Notes\\
	\hline
  8 Planets &  0.91183483 & 0.1416661 & Averages over 1000 \\
	          &               &            & random systems\\
	\hline
	7 Planets & 0.90784475 & 0.14490068 & idem\\
	\hline
	6 Planets & 0.90386207 & 0.15141145 & idem\\
	\hline
	5 Planets & 0.89719999 & 0.15552048 & idem\\
	\hline
	\end{tabular}
\caption{$R^2$ and $Median$ analysis for randomly created planetary systems. Maximum agreement $R^2_{max}=1$ and $Median_{max}= 0$ .}
\label{table4}
\end{table}

Already a superficial glimpse to Table \ref{table1} shows that the $R^2$'s of ``real" systems are much closer to 1 than the $R^2$'s attained by the randomly created systems of Table \ref{table4}. 
This first impression can be further substantiated by comparing ``real" and ``random" statistical indicators, as it is done in Table \ref{table5}.

 \begin{table}[!ht]
\centering
\begin{tabular}{|c|c|c|}
	\hline
	Systems &  AVG of all $R^2$'s  & AVG of all Medians \\
	\hline
  Randomly created &  0.90518541 & 0.148374678\\
	\hline
	Real & 0.966078109 & 0.06215032\\
	\hline
	\end{tabular}
\caption {Comparison between ``Randomly created" and ``Real" planetary systems, $R^2$ and $Median$ analysis.}
\label{table5}
\end{table}    

There, we compare the averages of the statistical indicators $R^2$ and $Median$ computed for artificial as well as for real planetary systems. From the figures, it appears extremely clear that an exponential distance relation (aka TB law) fits much better real planetary systems than artificial, randomly created, ones (see also the discussions in, e.g., Lecar (1973); Pletser (1988); Hayes \& Tremaine (1998); Pletser (2017)). Of course, the efficiency displayed by the TB rule in describing distances in real planetary systems calls for a (fully accepted and shared) theoretical explanation, which is perhaps still lacking, at present. However, from the phenomenological point of view, it is quite clear that the descriptive ability of the TB relation cannot be easily denied, even for extra solar planetary systems. 

One can also refine the ``quality" of the artificial planetary systems, for example by inserting mechanisms which mimic, in principle, the gravitational evolution of the (proto)planetary system itself. A possibility followed by Hayes \& Tremaine (1998) is to require that adjacent planets are separated by a minimum distance of $k$ times the sum of their Hill radii, for opportune values of $k$ ($0 \leq k \leq 8$). This originates more realistic planetary systems, and, according to the conclusion of Hayes \& Tremaine (1998), these systems generally fit a Titius-Bode law better than the purely ``random" ones, and about as well as some of the real planetary systems (as for example our own Solar System). From our point of view, this reinforces our thesis that the TB rule truly describes an actual phenomenological property of real planetary systems (Lara et al. 2012).\\

To illustrate further the capacity of the TB relation to effectively describe the spacing in planetary systems, we provide hereafter a diagram where the TBR-valid systems are related to a ``total number'' of observed exoplanetary systems.
Of course, crucial is the definition of what ``total number'' should mean in this context. Given our argued choice to restrict the investigation to systems with at least 5 planets, it would result into a nonsense to compare the number of TBR-valid systems with the totality of known planetary systems. In fact, we have sufficiently clarified that to include in our analysis also systems with 2 or 3 planets would produce meaningless results from the point of view of the semi-log linear regressions and the statistical indicators $R^2$ and $Median$. Obviously, there is only a single straight line passing through two points and in such 2-planets systems we would get $R^2=1$, by definition. Just adding one or two points (namely, planets) would not produce yet meaningful results from the statistical point of view. Things become physically (and statistically) significant when the examined planetary system contains at least 5 planets. So, the ``total number'' of exoplanet systems against which we compare the number of systems where TBR is satisfied (to certain degree) is the number of systems with at least 5 planets orbiting a single star, namely the 32 systems of Table \ref{table1}.   

\begin{figure}
\includegraphics[scale=0.38]{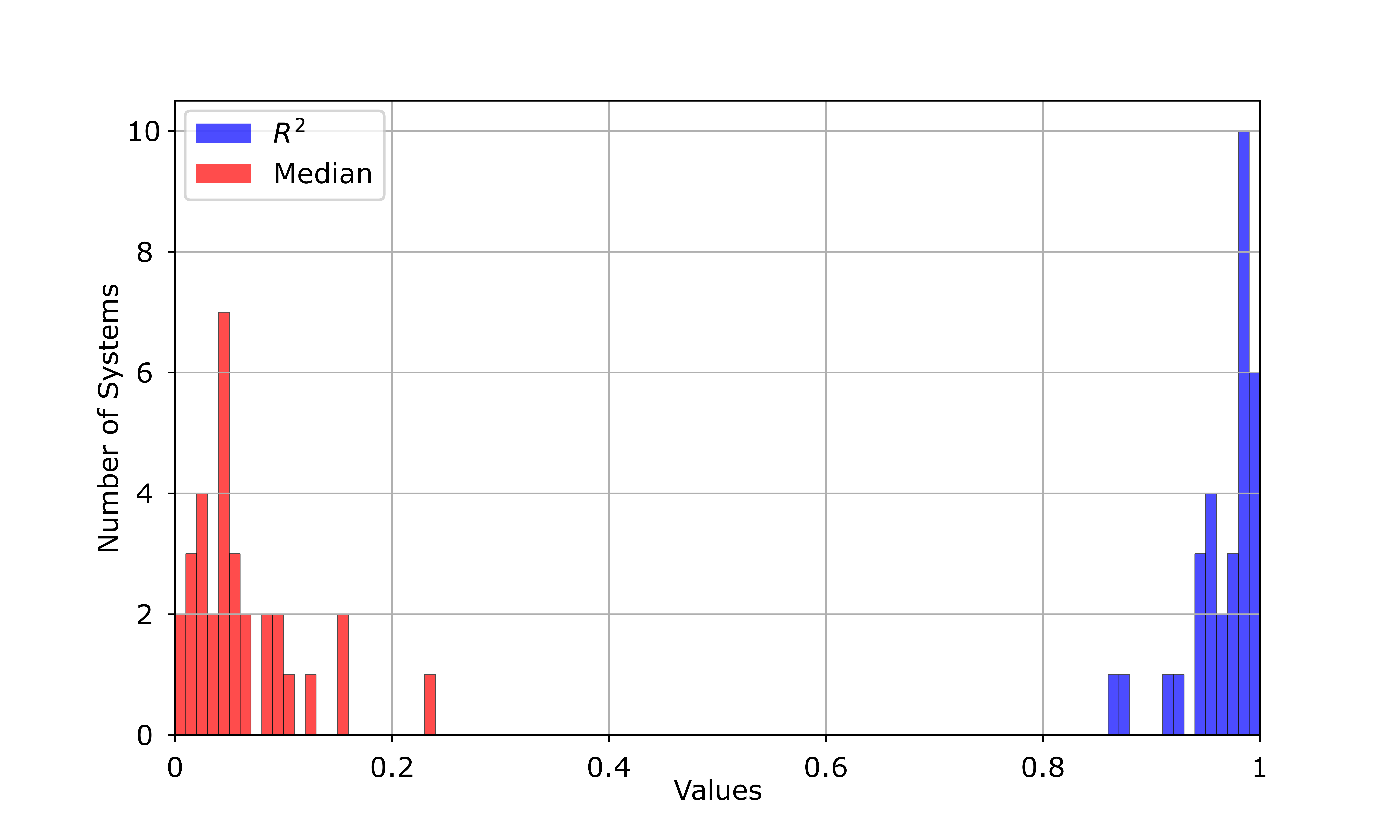}
\caption{Histogram of the $R^2$ and $Median$ values. Distributions of the $Median$-values (red, on the left) and of the $R^2$-values (blue, on the right) among the 32 exoplanetary systems listed in Table \ref{table1}. The bin size is $0.01$.}
\label{Histogram_R2_and_Median_Values.png}
\end{figure}

Having stated the above, in Fig.\ref{Histogram_R2_and_Median_Values.png} we show a histogram where we display the distribution of the $R^2$-values among the 32 systems considered in Table \ref{table1} (namely how many systems have $R^2>0.99$, how many with $0.98<R^2<0.99$, how many with $0.97<R^2<0.98$, etc., until the worst $R^2 \simeq 0.86$ exhibited by Kepler 150). The $R^2$ histogram is depicted in blue, while an analogue histogram describing the distribution of the 
$Median$-values is depicted in red.
From the histograms it is quite clear that the TB relation is very successful in mimicking the structure of real planetary systems. According to Table \ref{table1}, 6 systems out of 32 ($19\%$) have an $R^2>0.99$, and 16 out of 32 (50\%) have an $R^2>0.98$. Only 7 systems out of 32 (22\%) present an $R^2<0.95$ (and they will be specifically studied in the next Section, since they are very good candidates for the ``prediction'', or insertion, of new planets). These numbers speak by themselves in favor of the remarkable efficiency of the Titius-Bode relation for the description of orbital spacing in planetary systems with at least 5 planets each.

%

%
%
%

%
%
%
%
%
\section{Predicting new planets with the Titius-Bode rule}

From its conception, Titius-Bode rule was used as a predictive tool for discovering orbits of some celestial bodies in our Solar System. The original form of the law described well the distances of the known planets from the Sun, provided
a gap between Mars and Jupiter was admitted. Since the law was working so well for all the known planets, and moreover it adapted almost perfectly to the ``new comer" Uranus (discovered in 1781), people started taking it seriously, and searched a planet in between Mars and Jupiter for about twenty years. The endeavor was crowned with success when finally Piazzi observed Ceres (the first and largest asteroid) in 1801. 

In the previous sections we showed that the modern formulation of the law, Eq.~\eqref{tbl}, is in good agreement with the recent data of a large number of exoplanetary systems. Therefore it sounds natural to attempt harnessing TB rule, in order to successfully predict orbits of possible ``missing'' celestial bodies, perhaps just in those systems where the agreement with the law is, apparently, less striking (on this see also e.g. Bovaird \& Lineweaver (2013); Scholkmann (2013); Huang \& Bakos (2014); Bovaird et al. (2015); Lara et al. (2020)). 

Our procedure for achieving such predictions is described in what follows, and it appears quite straightforward. To our understanding, the predictions are readily experimentally testable in present or future exploratory exoplanetary missions (see e.g. PLATO; ARIEL; TOLIMAN). 

\begin{figure}
\includegraphics[width=\linewidth]{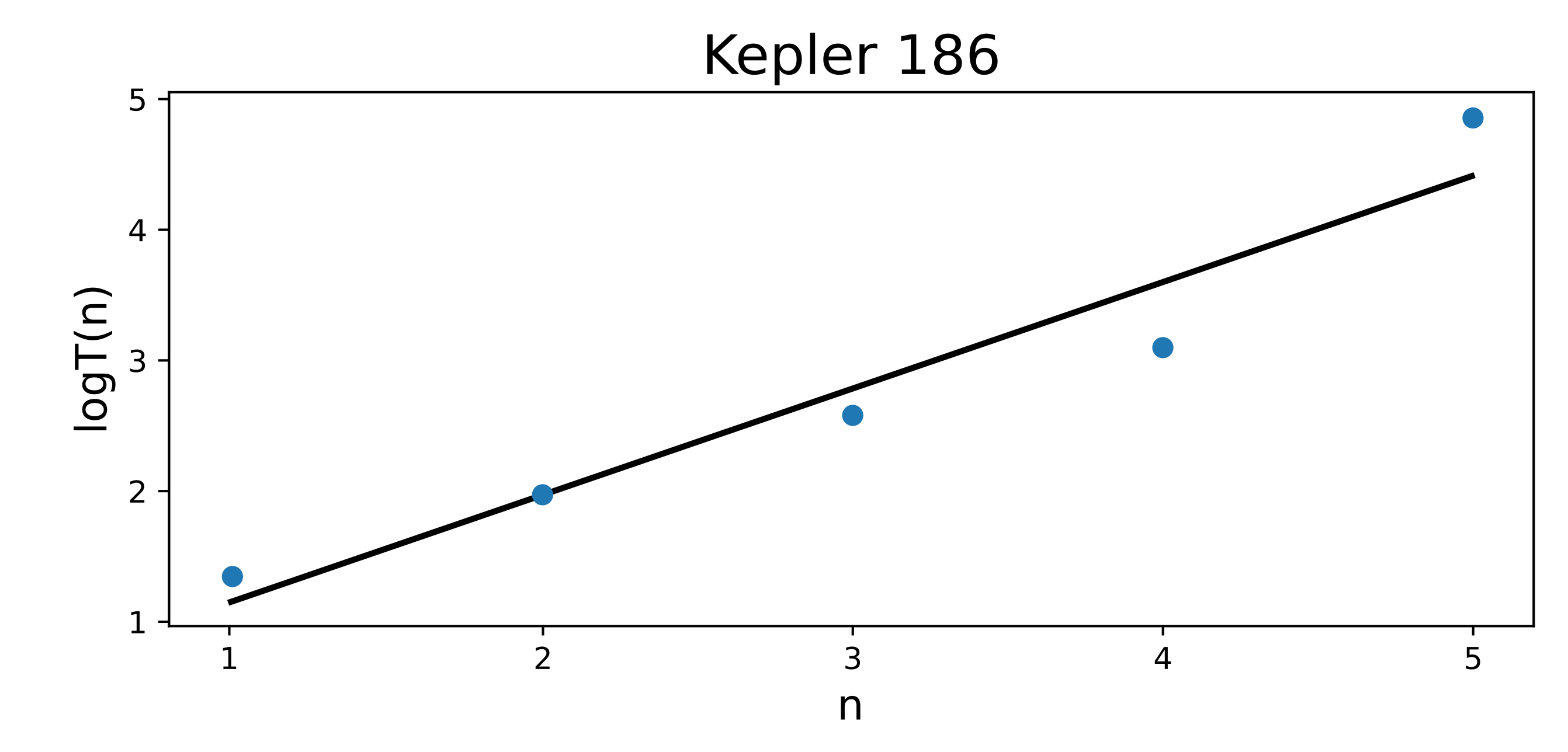}
\caption{This is a ``raw'' fitting of the Kepler 186 system, namely before the insertion of any gap. The statistical indicators (see Table~\ref{table1}) are $R^2=0.92681234$ and $Median = 0.093772029$. We can clearly identify a ``big'' jump between the fourth and the fifth planet.}
\label{Kepler186_corrected_raw.png}
\end{figure}

\begin{figure}
\includegraphics[width=\linewidth]{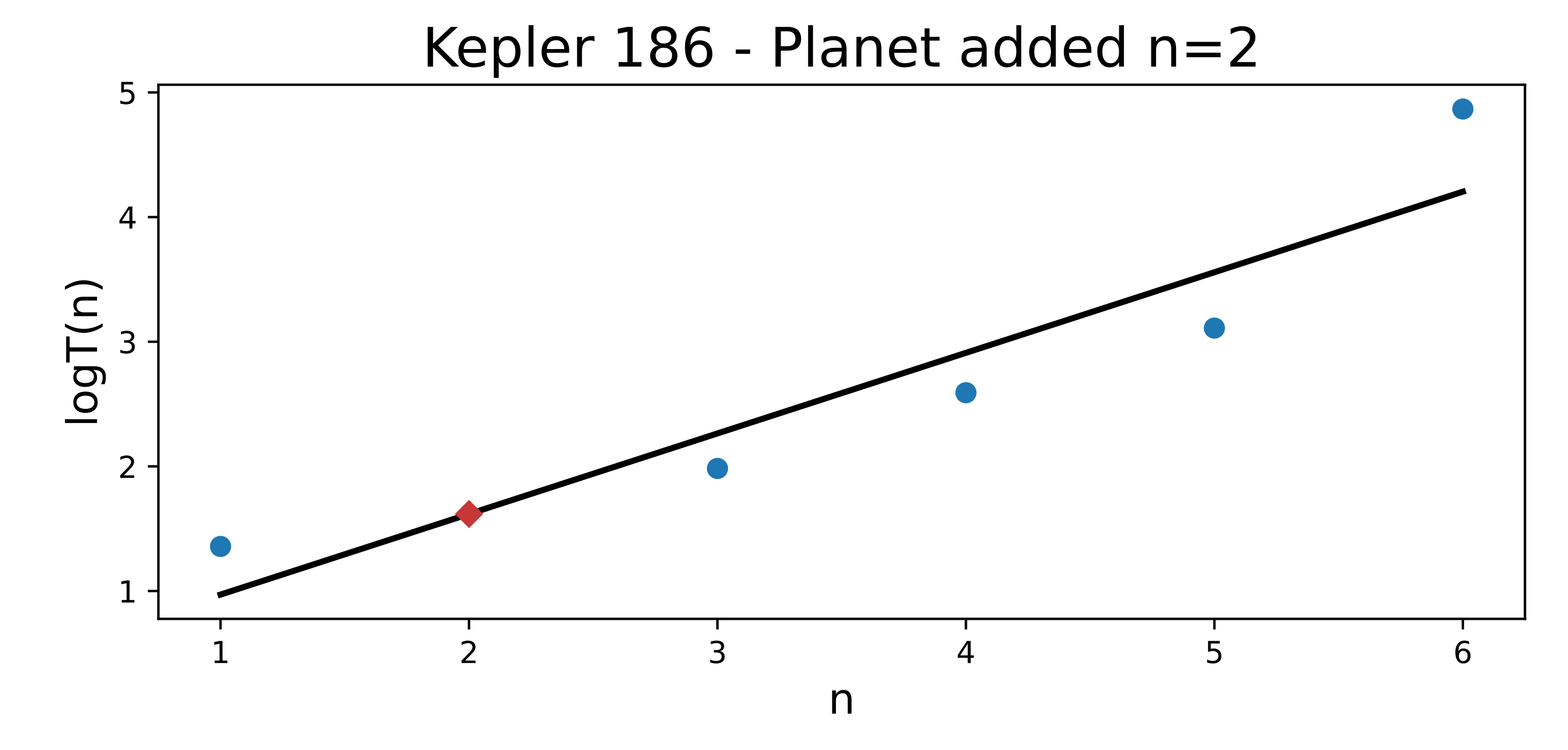}
\caption{In the Kepler 186 system, inserting for example a planet between the first and second planet makes the fit to TB law worse, i.e. $R^2$ decreases and the $Median$ increases significantly. The position of the hypothetical planet (represented by the red square) is $n=2$. [$R^2 = 0.864350834$, $Median = 0.141705383$]}
\label{FigKeplerWrong}
\end{figure}

\begin{figure}
\includegraphics[width=\linewidth]{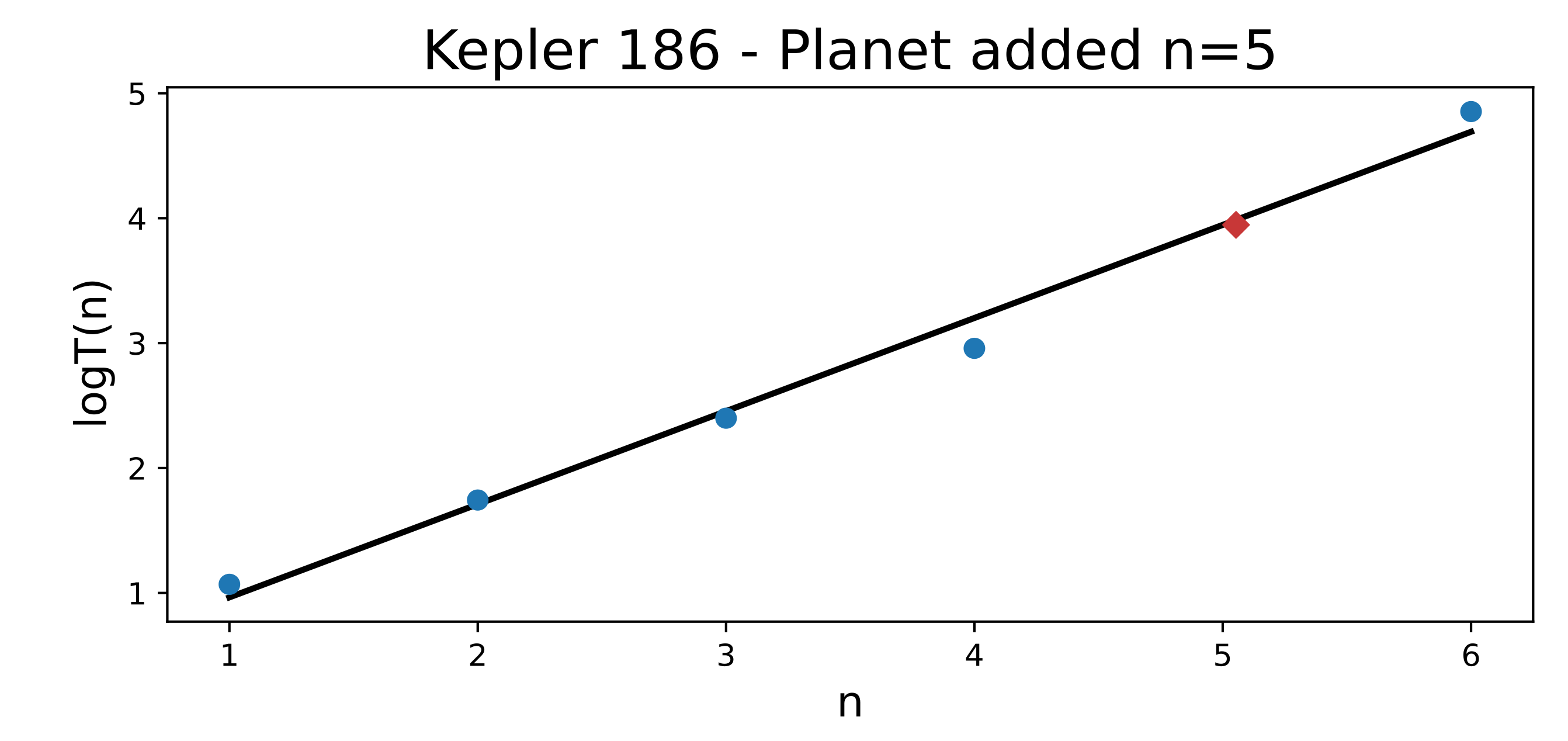}
\caption{In the Kepler 186 system, inserting a planet between the fourth and fifth planet makes the fit to TB law much better, i.e. improves $R^2$ and $Median$ significantly. The position of the hypothetical planet 
(represented by the red square) is $n=5$. [$R^2 = 0.98794486$, $Median = 0.030801858$]}
\label{FigKeplerCorrect}
\end{figure}

To start with, we give a quantitative definition of \textit{optimal agreement} for a TB fitting by using the $R^2$ coefficient of determination: any system with an $R^2 \geq 0.95$ is said to be in \textit{optimal agreement} with the TB rule. The idea now is that for the few systems where the agreement with TB rule is not \textit{optimal}, we can quite easily identify at least a (big) ``jump'' in the fitting graph (see e.g. the ``jumps'' in Fig.\ref{Kepler186_corrected_raw.png}). We then interpret the ``jump'' as a missing celestial object, whose orbital period (or position) can be predicted using the TB fitting. Since the values of  $\log T$ on the Y axis are fixed (namely given by the observations), then, in order  to ``smooth'' the jump in the graph (focus on Fig.\ref{Kepler186_corrected_raw.png} as an example), the only freedom we have is to insert a gap on the X axis, namely to shift the numerical places of the planets on the X axis.
Below, we discuss an example and a detailed explanation of our predictive procedure for the few exoplanetary systems with a non-optimal agreement. 
  
\begin{table}[!ht]
\centering
\begin{tabular}{|c|c|c|c|c|}
	\hline
	Systems & Old $R^2$ & New $R^2$ & Old Median & New Median \\
	\hline
	HD 219134 & .913657699 & .974791203 & .126773358 & .05855297 \\
	\hline
	HD 191939 & .945551901 & .985952278 & .08825776 & .034601236\\
	\hline
	K2-138 & .945458309 & .991472463 & .094321758 & .035982039\\
	\hline
	Kepler-150 & .8605413 & .952818595 & .236636167 & .152566663\\
	\hline
  Kepler 169 & .875354644 & .960693503 & .151151179 & .091297799\\
	\hline
	Kepler 186 & .92681234 & .98794486 & .093772029 & .030801858\\
	\hline
	Kepler 33 & .948841017 & .991408774 & .054322078 & .023090953\\
	\hline
	\end{tabular}
\caption{$R^2$ and $Median$ analysis before and after the insertion of one planet, for exoplanetary systems with not optimal agreement ($R^2 < 0.95$) according to Titius-Bode rule.}
\label{table6}
\end{table}


First, we collect in Table~\ref{table6} the planetary systems (chosen from those listed in Table~\ref{table1}), whose agreement is non-optimal (namely $R^2<0.95$).

We consider then, as a first example, Kepler 186, a system with 5 planets. We compute the $R^2$ resulting from a direct linear regression of the 5-planets system, without any gap inserted (Fig.\ref{Kepler186_corrected_raw.png}). Then we check whether the insertion of one gap improves the $R^2$ and $Median$ (significantly) or not. We note that if a gap (and therefore a planet) is placed at \textit{any} position other than that of the big ``jump'', the fit with TB becomes worse ($R^2$ worsens), as it can be seen from Fig.\ref{FigKeplerWrong}. Instead, if the gap is placed at the big ``jump" (between the fourth and the fifth planet), both $R^2$ and $Median$ are significantly improved, hence the fit with TB becomes much better (Fig.\ref{FigKeplerCorrect}). This point is very important: in most cases, the addition of a planet in an arbitrary place, even if the planet follows \textit{exactly} the TB law, makes the overall fit worse. Only when the planet is placed at the right point of a (big) ``jump'', then the fit becomes better; and significantly at that point.

Once the above procedure is applied to all the systems contained in Table~\ref{table6}, the experimentally testable outcome are \textit{the orbital periods of the predicted planets}, displayed in Table~\ref{table7}. We point out that, using this method, it would be possible in principle to insert more than one gap (even not consecutive), and therefore more than one planet, in a TB fitting. We refrain however to do that, since it would probably increase the bias of the data too much. For example, adding 2 points to a set of 5 would affect the $40\%$ of the total data, or even adding 2 points to a set of 7 would imply to bias almost the $30\%$ of the data. A further argument in favor of being sparing in the addition of new gaps or planets in a planetary system is the following: the geometrical intuition suggests that, given two bunches of points on a diagram, the $R^2$ of a linear regression among those points will be in general improved by ``moving away" one bunch of points from the other. So, in general, it looks wise not to abuse with the insertion of ``new" planets.
\begin{table}[!ht]
\centering
\begin{tabular}{|c|c|c|}
	\hline
	Systems & Position of       & Predicted \\
	        & predicted planet  & period (days) \\
	\hline
   HD 219134 & n=6 & 488.3976385\\
	\hline
	HD 191939 & n=6 & 744.2371442\\
	\hline
	K2-138 & n=6 & 23.05233175\\
	\hline
	Kepler 150 & n=5 & 145.3213414\\
	\hline
	Kepler 169 &  n=5  & 36.08528335\\
	\hline
	Kepler 186 & n=5 & 56.03297983\\
	\hline
  Kepler 33 & n=2 & 30.00633542 \\
	\hline
\end{tabular}
\caption{Here we display \textit{experimentally testable} predictions of the sequence positions and the periods of the planets ``predicted" with TB rule.}
\label{table7}
\end{table}
%
%
%
%
%
%
The possible reasons why these predicted missing planets have not yet been detected in previous surveys may be various, the first being, of course, the general difficulty of these measurements. A further realistic possibility is that the \textit{missing planets} are not really planets, but instead for example \textit{asteroid belts}, or clouds of dust (as it happens in our solar system). This possibility could be supported by the fact that, apart from Kepler 33, all the other systems allow for a gap at $n=5$ or $n=6$. Actually, this was the case of \textit{our} solar system, where the gap was exactly at $n=5$, between Mars and Jupiter. To showcase this possibility, we point out that the $R^2$ and $Median$ of a TB fitting of our solar system without the inclusion of Ceres or the asteroid belt are, respectively, $R^2_\text{old} = 0.981440834$ and $Median_\text{old} = 0.034928964$; to be compared with the current values, which include the gap at $n=5$ between Mars and Jupiter, i.e. $R^2_\text{new} = 0.993448869$ and $Median_\text{new} = 0.019581007$.

Finally, we would like to emphasize that the point of this section is \textit{not} to improve the agreement of the TB rule with systems of sub-optimal agreement. 
The \textit{sole} purpose of this section is to provide a testable prediction of the TB rule, which we believe may help in the search and discovery of new exoplanetary celestial objects.











%
%
\section{TB rule and the age of the planetary systems}

It is clear that TB rule is intended to be obeyed on a statistical basis, by planetary systems which are already ``running" by a ``fair" amount of time. At the beginning of their history, proto-planetary systems are obviously full of dust, small debris, rocks etc. which can be found at any distance from the central star. With time, the planetary formation process goes on, matter aggregates onto specific bodies, and planets slowly emerge. In particular, preferred orbits, those described for example by TB rule, slowly emerge and become more precise and definite. On the other hand, it is well known that light objects, like comets, small debris, in general small objects, do not (and are not expected to) obey, singularly, the TB rule, at any time. 
Therefore, it seems reasonable to investigate a possible correlation between the age of the planetary system, i.e. the age of the central star, and the goodness to which the TB rule is obeyed in that planetary system. Hereafter the reader will find such analysis.    

In Table \ref{table8} we report, like in Table \ref{table1}, the name of the planetary system (i.e. the name of the central star), the number of planets in the system, the $R^2$ indicator, but in place of \textit{Median} column we have the age of the central star in Gigayears. The age of the central star in general is not explicitly given on the NASA Exoplanet Archive (NASA Archive), however it can be obtained from literature dedicated to the specific planetary system under scrutiny.

\begin{figure}[h!]
  \includegraphics[width=8.5cm]{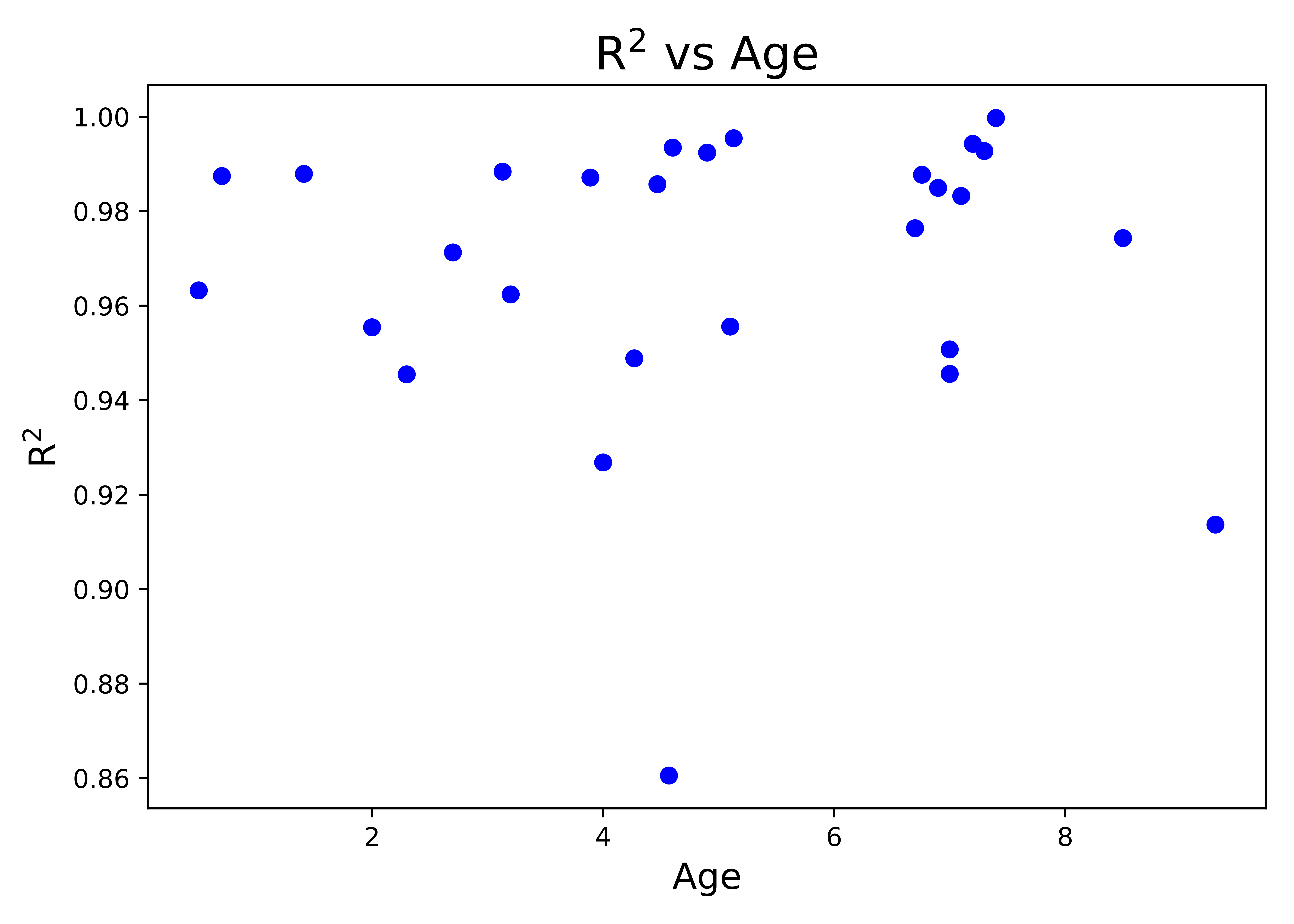}
 \caption{Diagram reporting $R^2$ vs Age of the Central Star.}
\label{AgeStar}
 \end{figure}

According to the above argument, in a diagram ``$R^2$ vs Age of Star", one would expect several points pretty far from 1 for young systems, and then for increasing ages, the points should accumulate nearby $R^2=1$. Instead, looking at Fig.\ref{AgeStar}, the situation is not such. We see young systems close to $R^2=1$, as well as older systems; and equally, young systems far from $R^2=1$, as well as old systems. Thus, according to the diagram in Fig.\ref{AgeStar}, no correlation emerges between the age of a planetary system, and the ``goodness" to which the system obeys TB rule. 
\begin{table}[!ht]
\centering
\begin{tabular}{|c|c|c|c|c|}
	\hline
	System & Planets & $ R^2 $ & Age of & Notes\\
	       &         &         & Star (Gyr) &       \\
	\hline
    Sun & 10 &  0.993448869  & 4.603 & with Ceres\\
	\hline
	KOI-351 &  8 & 0.96322542 & 0.5 & Kepler90\\
	\hline
	Trappist 1 & 7  & 0.994255648 & $7.2 \pm 2.2$ &\\
	\hline
	HD 10180 &  6  &  0.992725066 &	7.3 &\\
	\hline
	HD 191939 &  6  &  0.945551901  &	$7 \pm 3$  & 2022\\
	\hline
	HD 219134 &  6  &  0.913657699 &	$9.3$ &\\
	\hline
	HD 34445 & 6  &  0.974290137	& $8.5 \pm 2.0$ &\\
	\hline
	K2-138 &  6  &  0.945458309  & $2.3\pm^{0.44}_{0.36}$  & 2021\\
	\hline
	Kepler 11 &  6 & 0.962373014 &	$3.2 \pm 0.9$ &\\
	\hline
	Kepler 80 &  6 & 0.955414548	& 2 &\\
	\hline
	TOI - 1136 &  6 & 0.987422135 & 0.7 & 2022\\
	\hline
  TOI - 178 & 6 & 0.983217838  & $7.1 \pm 5.3$  & 2021\\
	\hline
	HD 108236 &  5 & 0.976381637	&  $6.7 \pm 4$  & 2021\\
	\hline
	HD 158259 &  5  & 0.999713447	&  7.4 & 2020\\
	\hline
	HD 23472 &  5  & 0.988496548	&   & 2022\\
	\hline
	HD 40307 &  5  &  0.984933377	& $6.9 \pm 4$ &\\
	\hline
	K2-268 & 5 &  0.957182918 &   & 2019\\
	\hline
	K2-384 & 5 &  0.983020508 &   & 2022\\
	\hline
	Kepler 102 & 5  &  0.987900552 & 1.41 &\\
	\hline
	Kepler 122 & 5 &  0.98710756 & 3.89  &\\
	\hline
	Kepler-150 & 5  & 0.8605413 &	4.57 &\\
	\hline
	Kepler 154 & 5 & 0.985703723 & 4.47 &\\
	\hline
	Kepler 169 &  5 &  0.875354644 &  &\\
	\hline
	Kepler 186 &  5 &  0.92681234 &	$4.0 \pm 0.6$ &\\
	\hline
	Kepler 238 & 5 & 0.98771232 & $6.76 \pm 2$ &\\
	\hline
	Kepler 292 & 5 &  0.995427692 & $5.13 \pm 3$ &\\
	\hline
	Kepler 32 &  5 &  0.971262925 &	$2.7 \pm 1$ &\\
	\hline
	Kepler 33 &  5 &  0.948841017	& 4.27 &\\
	\hline
  Kepler 55 & 5 & 0.988367826 & $3.13$  &\\
	\hline
  Kepler 62 & 5 &  0.950734271 & $7 \pm 4$ &\\
	\hline
	Kepler 82 & 5 & 0.955569351  & 5.1  & 2019\\
	\hline
Kepler 84 & 5 &  0.992394962	& 4.9 &\\
	\hline
\end{tabular}
\caption{$R^2$ and age of the central star (in Gyr). Data source: NASA Exoplanet Archive, plus literature on the specific planetary system considered.}
\label{table8}
\end{table}

\section{Comparison between the TB rule and the Harmonic Resonances method}
Of course, the TB rule does not represent the only attempt to give a phenomenological description of planetary spacing. One of the most interesting and promising ways to arrive to such a description is the so called \textit{Harmonic Resonances} (HR) method. The idea comes from the observation that points of a periodic system, interacting with each other (as for example the  orbital systems), tend to manifest resonances, and resonances tend to stabilize the periodic system itself. In particular, computer simulations of planetary systems have shown that random injections of new planets tends to produce dynamically unstable orbits, until the planets surviving in the system settle their orbital periods into ``harmonic ratios'', namely in (quasi) rational ratios, usually expressible in simple fractions (on this see Peale (1976); McFadden et al. (2007); Batygin, Morbidelli (2013); Aschwanden (2018)).       
Historically, the HR idea seems to originate from a generalization of a resonance relation already known to Laplace (1829) among the orbital periods of three Galileian satellites of Jupiter, namely Io, Europa, Ganymede, which can be written as 
\be
\frac{1}{T_{Io}} - \frac{3}{T_{Eur.}} + \frac{2}{T_{Gan.}} \simeq 0 \,.
\label{Jupiter}
\ee   
Such relation is fulfilled with an accuracy of order $\approx 10^{-5}$, and, using more precise orbital periods, perhaps even to an astonishing accuracy of $\approx 10^{-9}$ (Peale 1976). 

On the ground of the orbital periods known in the Solar System, Aschwanden (2018) proposed a generalization of Eq.\eqref{Jupiter} for a two-body resonance existing between \textit{two neighbored planets} in stable long-term orbits, as
\be
\frac{H_i}{T_i} - \frac{H_{i+1}}{T_{i+1}} \simeq \omega_{i, i+1}\,.
\label{2body}
\ee
The $H_i$ are (small) integers, $T_i$ is the orbital period of the $i$-th planet, and $\omega_{i, i+1}$ is the residual that accounts for possible further resonances from third or more planets involved. As Aschwanden explains, once considered two neighbored planets, we know from observation $T_i, T_{i+1}$, and we choose the two smallest integers $H_i, H_{i+1}$ (positive or negative) such that the modulus of the residual $|\omega_{i, i+1}|$ results to be minimal. The ratios $H_{i+1}/H_i$ are called \textit{harmonic ratios}. Of course, once a particular ratio between two periods is established, $T_{i+1}/T_i \simeq H_{i+1}/H_i$, this can be immediately transferred into a ratio between the distances of the planets from the central body, by virtue of Kepler's third law, $R \propto T^{2/3}$. 
On the basis of previous empirical work (Peale 1976), the integers $H_i$ are picked in the range $1,...,5$, and therefore the possible distinct harmonic ratios considered are $(5:4), (5:3), (5:2), (5:1), (4:3), (4:1), (3:2), (3:1), (2:1)$. This procedure is applied by Aschwanden to all the 9 neighbored planet pairs that can be extracted from the Solar System, including Ceres (considered here as the main representative of the asteroids belt). So the pairs considered are: Mercury-Venus, Venus-Earth, Earth-Mars, Mars-Ceres, Ceres-Jupiter, Jupiter-Saturn, Saturn-Uranus, Uranus-Neptune, Neptune-Pluto. Empirically, Aschwanden finds for the above 9 planet pairs that the the best-fit resonances are confined to the five particular ratios $(3:2), (5:3), (2:1), (5:2), (3:1)$, and the residuals are all quite small, being in the range $\omega_{i, i+1}\approx 0.005 - 0.06$. Moreover, it is found that the ratio $(5:2)$ works for four different pairs Mercury-Venus, Mars-Ceres, Ceres-Jupiter, Jupiter-Saturn, the ratio $(2:1)$ describes the pairs Earth-Mars and Uranus-Neptune, while the remaining three ratios fit one single pair each.

Also the role of the most massive planet, Jupiter, is investigated (Aschwanden 2018) by expanding the previous 2-body equation \eqref{2body} into a 3-body resonance condition
\be
\frac{H_i}{T_i} - \frac{H_{i+1}}{T_{i+1}} - \frac{H_{Jup}}{T_{Jup}} \simeq \omega_{i, i+1} \,.
\label{3body}
\ee
However, identical results were found (yielding $H_{Jup}=0$), except for the 3-body configuration Mars-Ceres-Jupiter. This supports the conclusion that the neighbored planet-planet interaction is more important in shaping the resonance than the influence of the largest giant planet, with the exception perhaps of the planet-asteroid pairs. 

Once, through empirical attempts, the harmonic ratios are established, they can be used to re-construct distances among planets in the Solar System, of course via the Kepler third law, namely
\be
\frac{R_{i+1}}{R_i} = \left(\frac{T_{i+1}}{T_i}\right)^{2/3} \simeq \left(\frac{H_{i+1}}{H_i}\right)^{2/3} \,.
\label{RT}
\ee 
This procedure is applied by Aschwanden to the planets of the Solar System, as well as to 7 moons of Jupiter, to 13 moons of Saturn, to 8 moons of Uranus, and to 6 moons of Neptune. All the satellites are chosen according to the rule of having a diameter larger than 100 km, on the ground of the general rule that planetary spacing patterns (as TB rule, or HR rule) are in general obeyed by ``enough big" objects. It is in fact well-known that ``small" objects, like comets, small asteroids, debris, etc. do not follow, singularly, any particular pattern in the size of their orbits, and can be found at any distance (allowed by classical mechanics) from the central body. 

For all the above planetary systems,  HR model produces an agreement between observed planetary distances $R_{obs}$ and model-predicted values $R_{pred}$ significantly better than the Titius-Bode law \eqref{tbl}. The reported quantities by Aschwanden (2018) are the \textit{mean} and the \textit{standard deviation} of the ratios $R_{pred}/R_{obs}$ computed for the satellites of a given planetary system. For example, for the Solar System the values for the TB law are $R_{TB}/R_{obs}=0.95 \pm 0.13$, to be compared against a $R_{HR}/R_{obs}=1.00 \pm 0.04$ given by the HR model. Similar good figures are described in (Aschwanden 2018) for the satellite systems of Jupiter, Saturn, Uranus, Neptune, and even for two extra solar planetary systems, namely 55 Cnc, and HD 10180. Thus, Aschwanden (2018) concludes that the HR model is clearly better than the TB rule in describing planetary spacing, and can be also much more efficiently used to ``predict" possible missing planets or satellites in ``empty" places of the planetary sequences (Aschwanden \& Scholkmann 2017; Scholkmann 2017)
\footnote{At a first sight, one further possibility to compare HR model and TB rule could be to consider a $R^2$-test for the HR method applied to Solar System, so to compare it with the $R^2$-test applied to the TB rule, also for the Solar System. However, $R^2$-tests are well defined, in general, for predictive models encoded by a function, typically a linear regression, maybe coming from exponential rules through taking a log (see Eqs.\eqref{tbl},\eqref{logTB}). On the contrary, from Eq.\eqref{RT} it is clear that the HR method does not generate an exponential law, neither a definite function. This would happen if the ratios $H_{i+1}/H_i$ were the same for all the planet pairs, say $(H_{i+1}/H_i)^{2/3} = \alpha$. Then from Eq.\eqref{RT} one would infer $R_{i+1}=\alpha R_i$, and hence $R_n = \alpha^{n-1} R_1$, which is the TB rule. But in general the ratios $H_{i+1}/H_i$ are not the same for different pairs, and actually they change from pair to pair in an (in principle) unpredictable way. Therefore the $R^2$-test seems to be non-applicable to the HR rule. So, in order to compare TB and HR rules we have to stick on the Aschwanden method $R_{pred}/R_{obs}$, above already described.}.

Although the above figures should not certainly be under-appreciated, some considerations on the comparison between the TB rule and the HR method are surely in order. HR method seems to have some grounds in the classical mechanics of periodic systems, perhaps firmer and more sounding than the grounds backing the TB rule.  However, as a matter of fact, there is no definite theory, or bunch of theorems, in classical or celestial mechanics predicting that a periodic system should exhibit (harmonic) resonances, after that a certain amount of ``running" time has elapsed. And even less, no theory predicts what \textit{specific} harmonic ratios should be used to describe a particular system. The integers entering the ratios in  Eqs.\eqref{Jupiter},\eqref{2body},\eqref{3body},\eqref{RT} are chosen empirically, \textit{ad hoc} for any specific planetary pair, among ``small" integers. Of course, being a matter of rational numbers, if the ratio (5:2) describes well, say, the pair Jupiter-Saturn, nothing prevents us from thinking that perhaps the ratios (7:3), or (8:3), or maybe (51:22), or (48:19), can describe that pair even better. In other words, the arbitrariness in the choice of a given ratio is large. A further weak point of the HR method is that no theory predicts which ratio should be used for a particular planetary pair, or if/why ratios have to be repeatedly used (and how many times) to describe different pair of planets or satellites, as happens for example in the Solar System with the pairs Mercury-Venus, Mars-Ceres, Ceres-Jupiter, Jupiter-Saturn (and similarly in other (exo)planetary systems). The lack of any explicit theoretical constraint for choosing the numbers $H_i$, makes the HR method equivalent to a model with 5 or even 9 pairs of free parameters (namely, 10 to 18 free parameters in total). Then it comes with no surprise that the HR model fits the planetary distances (in our solar system, as well as in exoplanetary systems) with much higher accuracy than the TB law. Using a model that in principle has (at least) ten free parameters produces quite obviously better fittings than a model which has only two free parameters, as the TB rule in Eq.\eqref{tbl}. As said above, Aschwanden claims to have singled 5 harmonic ratios out of 9 which are particularly useful to describe planetary and satellite distances, both in the Solar and extra Solar systems. But, still, this is just an empirical, though extremely useful, observation. No reasons are given about why just those five ratios work well, or why the ratio (5:3) works better or worse than the ratio, say, 
(49:31). 

By comparison, TB rule has only two free parameters, that we indicate with $a$ and $\lambda$. The two free parameters have to be adapted (usually with a regression method) to the specific planetary system under consideration, and in general they change from system to system. The 5 harmonic ratios above proposed by Aschwanden, on the contrary, seem to be valid (again on an empirical basis) for any planetary system. However also the HR ratios have to be chosen each time \textit{ad hoc} for the specific system examined, and often with \textit{a priori} unpredictable repetitions.

All in all, we conclude that the TB rule still appears to be the empirical, \textit{most economic} and efficient rule on the ``market" to describe planetary distances or periods.

%
%
%
%
%

%

\section{In Laskar's footsteps: second degree polynomial \& Exponential}


Laskar (2000) presented an interesting model of planetary accretion, and consequent formation of a planetary system, based on the conservation of mass, momentum, and angular-momentum-deficit stability. Without entering in technical details, we can say that the model predicts the distribution of the final stable orbits of the planetary system as a function of the initial linear mass density $\rho(r)$ in the proto-planetary disk. For a mass density of the form $\rho(r)=\zeta r^p$, the distributions of the stable final orbits are given in Table \ref{table9}, for some specific values of $p$.    

\begin{table}[!ht]
\centering
\begin{tabular}{|c|c|}
	\hline
	$\quad p\quad$ & $r(n)$ \\
	\hline
  $\quad 0\quad$ &  $\quad r^{1/2} \ = \ A_1 \ + \ B_1 \ n \quad$  \\
	$\quad -1/2 \quad$ & $\quad r^{1/3} \ = \ A_2 \ + \ B_2 \ n \quad $ \\
	$\quad -1\quad$ & $\quad r^{1/6} \ = \ A_3 \ + \ B_3 \ n \quad $ \\
 $\quad -3/2 \quad$ & $\quad \log(r) \ = \ A_4 \ + \ B_4 \ n \quad$ \\
	\hline
	\end{tabular}
\caption{Planetary distributions corresponding to different initial mass densities $\rho(r)=\zeta r^p$ of the proto-planetary disks. A constant distribution $\rho(r)=\zeta$, i.e. $p=0$, gives a law in $n^2$; while $p=-3/2$ gives a Bode-like exponential law. Of course, $n=0,1,2,3,\dots$.}
\label{table9}
\end{table}

The parameters $A_i$, $B_i$ are in general functions of the angular momentum deficit, and of the (arbitrary) parameter $\zeta$ (see (Laskar 2000) for more details). In the context of the present work, the parameters $A_i$, $B_i$ can be determined from the observational data via the usual linear regression method. We notice that all the planetary distribution laws predicted by the Laskar's model contain \textit{only}  two (free) parameters $A_i$, $B_i$. From this point of view the TB rule and the other distributions proposed in Table \ref{table9} are similar. On the contrary, they differ blatantly from the HR rule, where the potentially free parameters are at least five or ten. 

Particularly significant is the law with $r \sim n^2$. In fact, power laws as $n^2$ were proposed in several works, even long ago, for the distribution of planets or satellites in the Solar System (see Caswell 1929; Schmidt 1945; Nottale 1996; Nottale et al. 1997; Agnese \& Festa 1997; Reinisch 1998; de Oliveira Neto et al. 2004). 
\footnote{It is interesting to observe that, with the exception of Schmidt, all the authors in (Caswell 1929; Schmidt 1945; Nottale 1996; Nottale et al. 1997; Agnese \& Festa 1997; Reinisch 1998; de Oliveira Neto et al. 2004) arrive to a quadratic law for the orbital radii distribution, $r \sim n^2$, because they consider a quantum-like description of the planetary systems, where they assume a Newtonian potential $1/r$ together with a Bohr-like quantization condition for the angular momentum (per unit mass). A different example of quantum-like description, that instead leads to an exponential rule $r\sim e^{2\lambda n}$, is presented in (Scardigli 2007).}
Although at a first glance the $n^2$ rule seems to be able to fit some data, more attentive considerations reveal that authors in (Caswell 1929; Schmidt 1945; Nottale 1996; Nottale et al. 1997; Agnese \& Festa 1997; Reinisch 1998; de Oliveira Neto et al. 2004) are in general forced to use non consecutive integers in order to accommodate the planetary distances. Or, perhaps more often, they split the Solar System into two sets of inner and outer planets (i.e. terrestrial and gigantic planets); the semimajor axis in each set then follow a $n^2$ power law to a high degree of approximation, however with different coefficients $A_i$, $B_i$ for each of the two planetary families. In other words, we have two different fitting parabolae, which in general do not join into a single smooth function.       

%
%
%
However, the sounding theoretical basis of the $n^2$ rule (at least, in the Laskar formulation), and its partial efficiency in describing planetary distances, have pushed us to check such law also in extra solar planetary systems.
Therefore we took the worst 6 systems in terms of $R^2$ and $Median$ agreement with the TB rule, and we recomputed them with a polynomial $(A+Bn)^2$ fitting, to see if their new $R^2$ and $Median$ indicators are better or worse than those obtained with the TB fitting. 

Since, according to Table \ref{table9} (Laskar), we have 
\be
r \sim (A+Bn)^2\,, 
\ee
and also, according to Kepler third law, $r^3 \sim T^2$, then it should hold 
\be
T^{1/3} \sim A+Bn\,. 
\ee
This last relation is the one we checked against the observational data. The ``goodness" of the linear regressions in getting the coefficients $A,B$ is displayed in Table \ref{table10}, as usual in terms of the $R^2$ statistical indicator.
The $R^2$ of the six worst TB systems (central column) are compared with the $R^2$ of the same systems, but re-computed via a $n^2$ fitting (right column). Just a simple glance is sufficient to conclude that the TB fitting, even in the worst six cases, works way better than the $n^2$ fitting. The TB rule seems to emerge, once again, as the most efficient, and most economical mathematical relation to describe planetary spacing, not only in the Solar System, but also in extra solar planetary systems.

For sake of completeness, we could perhaps check the $R^2$ of the worst 6 systems also against a model with $r \sim n^6$, listed also in the Laskar Table \ref{table9}. However, it is quite clear that the larger is the exponent of the power law, $r\sim n^k$, the closer such law is to the exponential $r \sim e^n$, namely to the TB rule. Therefore, such further fitting and comparison with the TB rule probably wouldn't result to be particularly insightful.

\begin{table}[h]
\centering
\begin{tabular}{|c|c|c|}
	\hline
	Worst TB Systems &  $R^2$ TB & $R^2$ Laskar $n^2$ \\
	\hline
	Kepler 150 & 0.8605413 & 0.696871005 \\
	\hline
  Kepler 169  &  0.875354644 & 0.769928531 \\
	\hline
	HD 219134 & 0.913657699 &	0.684612854 \\
	\hline
	Keper 186 & 0.92681234 &	0.819327821 \\
	\hline
	K2-138  & 0.945458309 & 0.860764702  \\
	\hline
	HD 191939  &  0.945551901 & 0.785594964 \\
	\hline
	\end{tabular}
\caption {$R^2$ of the six worst TB systems compared with the $R^2$ of the same systems, but re-computed via a $n^2$ fitting.}
\label{table10}
\end{table}

%

%

\begin{figure}[h!]
\centering
\includegraphics[width=8cm]{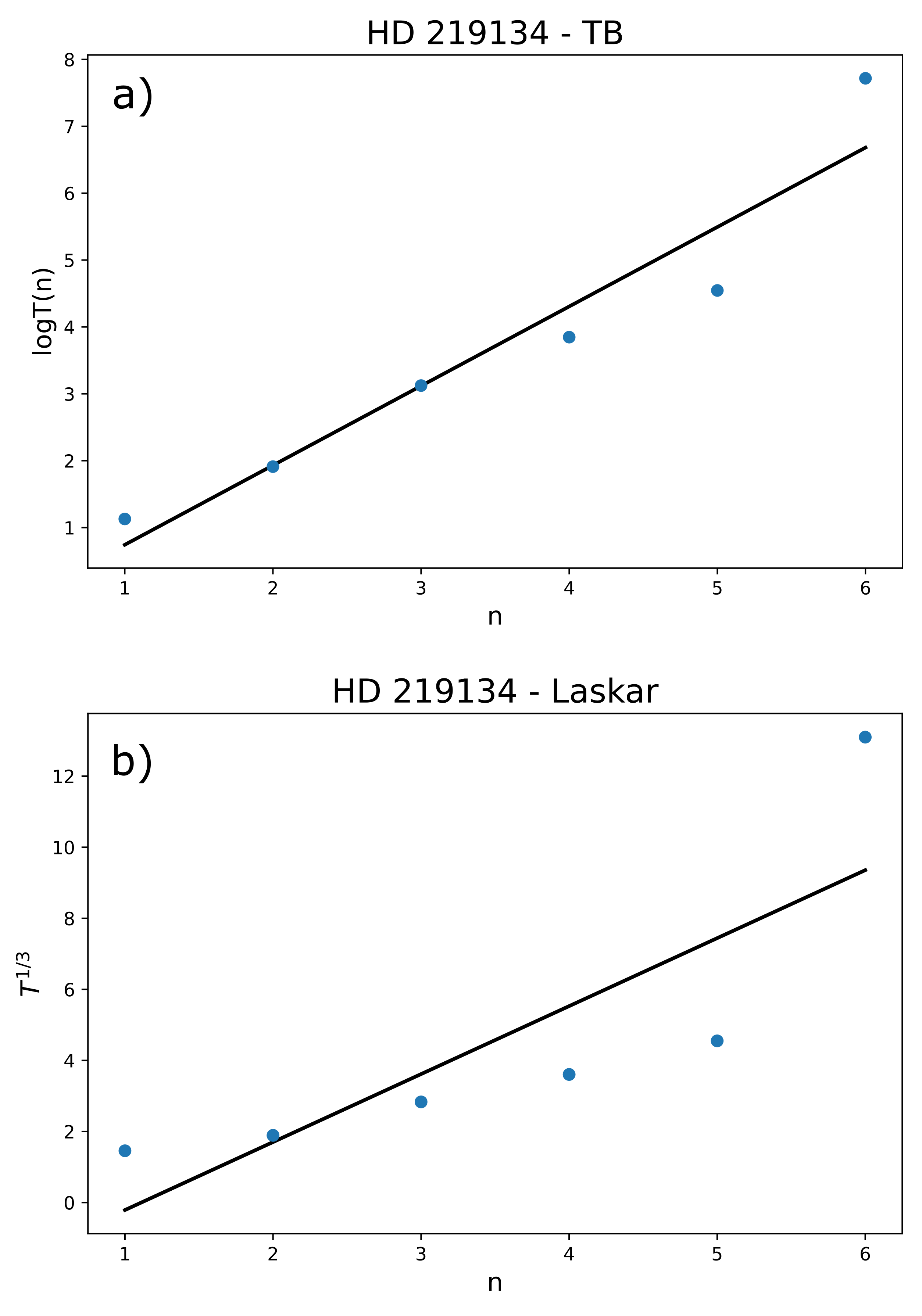}
\caption{Graphic comparison of two different fittings for HD 219134, a system with 6 confirmed planets.\\
\textit{a}) Titius-Bode fitting (namely semi-log) of HD 219134. The $R^2$ statistical indicator (Table~\ref{table10}) is $R^2=0.913657699$.\\
\textit{b}) $n^2$-fitting (Laskar model) of HD 219134. The $R^2$ statistical indicator is $R^2= 0.684612854$.}
\label{Laskar}
\end{figure}

Finally, for a visual comparison, we consider one of the systems in Table X, HD 219134, and we draw 2 diagrams of the same system: one displaying a TB-fitting and the other displaying a $n^2$-fitting, see Fig.\ref{Laskar}. Then, it appears extremely clear that a TB rule works way better than a $n^2$-fitting (Laskar model).

%
%

\section{Discussion and conclusions}
%
The aim of this paper is to check if the Titius-Bode rule can be considered a reliable description of the spacing among planets in a planetary system, with reference not only to our Solar system, but also in extra solar planetary systems. To this scope, after a general introduction to the Titius-Bode rule, we examined the 32 planetary systems (31 plus the Solar system) known to harbor at least 5 planets each (to date, July 2023). To keep the analysis as uniform as possible, we considered systems with a single central star only. We then computed the semi-log linear regressions, and used the statistical indicators $R^2$ and \textit{Median} in order to measure how ``good'' is an exponential distance relation, like the TB rule, in fitting the observational data. To clear the stage from the suspect of a purely random origin of the TB law, we created at random 4000 ``artificial'' planetary systems and observed that, according to the statistical indicators, the TB rule fits real planetary systems much better than artificial ones. We found that the TB rule describes more than $78\%$ (25 out of 32) of the examined real systems with an accuracy better than $R^2\geq 0.95$, and even more than $50\%$ with an $R^2\geq 0.98$. By comparison, the ``artificial'' random systems are fitted, on average, with an accuracy of $R^2\simeq 0.90$. 

Following a long tradition, we too use the TB rule to predict some new planets in systems that apparently present bad fittings. Our criterion to insert new planet(s) into a system is simple: the insertion should improve the fitting, namely the $R^2$ should getting closer to 1, and possibly the $Median$ getting closer to 0.  However, we also critically discuss the intrinsic limits that such a procedure seems to present naturally, and therefore the reasons of why it looks wise not to exceed in planetary insertions. A possible correlation between the age of a planetary system and the goodness to which TB rule is obeyed (which would seem a natural link), has been on the contrary ruled out by the observations. 

Our findings can be compared and contrasted with results of previous relevant works on the TB rule. Hayes and Tremaine (1998), for example, have considered, like us, randomly created planetary systems, but furthermore they selected specific systems by applying a conservative stability criterion which requires that adjacent planets are separated by a minimum distance of $k$ times the sum of their Hill radii (for values of $k$ ranging from 0 to 8). They performed least-squares fits of these systems to generalized Bode laws and compared them with the fit of our own Solar System. In so doing they found that this stability criterion generally produce geometrically spaced planets that fit a Titius-Bode law about as well as our own Solar System. From our point of view, this means that TB rule, far from being a mere product of chance, actually implies some kind of underlying physical mechanism at work. Thus, our findings which strongly support the validity of TB relation also in extra solar planetary systems, coherently fit with the results of Hayes and Tremaine, in pointing towards a definite universal physical law. 

On a different line, Bovaird and Lineweaver (2013), building (also) on the early data collected by the Kepler mission,   presented one of the first systematic studies on a possible validity of the TBR in extra solar planetary systems. They considered the 71 systems known in 2013 for harboring at least 4 planets each. It should be said, however, that some of these systems contain double or triple stars at their center, a circumstance that we were able to avoid in the present paper, thanks of course to the much richer set of exo-systems available to us, ten years later (2023). As statistical tool they employed a $\chi^2$ analysis, which however, in case of uniform or constant error bars (a position adopted by Bovaird and Lineweaver), is essentially equivalent to the $R^2$ analysis used in the present work (see Appendix A). These authors found that a vast majority of exoplanetary systems of their set adhere to the TB relation to an extent similar or greater than our Solar system does. This finding is essentially confirmed by our work, ten years later, and perhaps even reinforced, given the more refined set of planetary systems we used (five or more planets for each system, and single-star systems only). Then Bovaird and Lineweaver, on the basis of the just verified TB law, dive into a whirlwind of predictions of new planets. They arrive at inserting up to 10 new planets for each system, compiling a list for the existence of 141 new exoplanets in 68 multiple-exoplanet systems. On this point, in Sec.III we have been much more prudent (for the geometrical reasons we explained there), and we ``predicted'' at most one new planet per system. The criterium we adopted was however coincident with that of Bovaird and Lineweaver, namely the insertion of a new planet should improve the $R^2$ (or $\chi^2$) distribution.     

Finally, in Sections V and VI we compared the TB rule with two of its major competitors, namely the Harmonic Resonances (HR) method (see e.g. Aschwanden 2018), and the polynomial fitting (see e.g. Laskar 2000). In both cases, although for different reasons, the TB rule, an exponential relation,  emerges as the most ``economical'' (in terms of free parameters) and best fitting law for the description of the spacing among planetary orbits. 

In conclusion, we believe that our study contributes to definitively lift the TB rule out of the bag of conjectures and dubious numerical coincidences, and gives to the rule the status of a corroborated physical phenomenological law. Perhaps a bit like what the studies of Fraunhofer and Balmer did for the atomic spectra along the nineteenth century (for potential connections of TBR with atomic physics see Caswell (1929); Albeverio et al. (1983); Agnese et al. (1997); Scardigli (2007); Batygin (2018)). The  theoretical interpretations and explanations of such an empirical law still remain a much debated matter of the present-day research.


%
%
%
%
%

\section*{Data Availability Statement}

The data underlying this article are available in the NASA Exoplanet Archive, at\\ 
https://exoplanetarchive.ipac.caltech.edu/cgi-bin/TblView/nph-tblView?app=ExoTbls\&config=PS\\
For each planetary system displayed in Table \ref{table1}, the data are given in the specific references listed in Table \ref{table14}.


\section*{Acknowledgements}

We dedicate this work to the cherished memory of Jan Zaanen.
We thank Dr. Aravindh Swaminathan Shankar for his invaluable help with a Python code. F.S. also thanks G.'t Hooft and M.Arpino for useful discussions on the topic.

\section*{REFERENCES}
\small{
\noindent \textbf{Agnese A.G., Festa R., 1997,} Phys. Lett. A 227, 165

\noindent \textbf{Albeverio S., Blanchard P., H{\o}egh-Krohn R., 1983,} Exp. Math. 4, 365

\noindent \textbf{Alfven H., 1954,} On the Origin of the Solar System (Clarendon Press, Oxford)

\noindent \textbf{ARIEL} https://en.wikipedia.org/wiki/ARIEL

\noindent \textbf{Aschwanden M.J., Scholkmann F., 2017,} Galaxies, 5, 56

\noindent \textbf{Aschwanden M.J., 2018,} New Astronomy 58, 107

\noindent \textbf{Batygin K., Morbidelli A., 2013,} A\&A, 556, A28

\noindent \textbf{Batygin K., 2018,} MNRAS, 475, 5070


\noindent \textbf{Blagg M.A., 1913,} MNRAS 73, 414

\noindent \textbf{Bode J.E., 1772,} Anleitung zur Kenntniss des gestirnten Himmels, 2nd ed. (Hamburg), p. 462

\noindent \textbf{Bovaird T., Lineweaver C.H., 2013,} MNRAS, 435 (2), 1126

\noindent \textbf{Bovaird T., Lineweaver C.H., Jacobsen S.K., 2015,} MNRAS, 448 (4), 3608

\noindent \textbf{Caswell A.E., 1929,} Science 69, 384

\noindent \textbf{Chang H.Y., 2008,} J. Astron. Space Sci., 25, 239

\noindent \textbf{Chang H.Y., 2010,} J. Astron. Space Sci., 27, 1

\noindent \textbf{Cuntz M., 2012,} Publ. Astron. Soc. Japan, 64, 73

\noindent \textbf{de Oliveira Neto M., Maia L.A., Carneiro S., 2004,} Chaos, Solitons, Fractals, 21, 21

\noindent \textbf{Dermott S.F., 1968,} MNRAS, 141 (3), 349 

\noindent \textbf{Graner F., Dubrulle B., 1994,} A\&A, 282, 262

\noindent \textbf{Hayes W., Tremaine S., 1998,} ICARUS 135, 549

\noindent \textbf{Huang C.X., Bakos G., 2014,} MNRAS, 442, 674

\noindent \textbf{Jupiter moons} https://en.wikipedia.org/wiki/Jupiter\#Moons

\noindent \textbf{Kepler J., 1596,} Mysterium Cosmographicum (Georgius Gruppenbachius, Tubingen)

\noindent \textbf{Laplace, P.S., 1829,} Mechanique C\'eleste, Vols. I-IV. Hillard, Gray, Little and Wilkins., Boston

\noindent \textbf{Lara P., Poveda A., Allen C., 2012,} AIP Conference Proceedings 1479, 2356

\noindent \textbf{Lara P., Cordero-Tercero G., Allen C., 2020,} Publ. Astron. Soc. Japan, 72 (2), 24 (1-16)

\noindent \textbf{Laskar J., 2000,} Phys. Rev. Lett. 84, 3240

\noindent \textbf{Lecar M., 1973,} Nature 242, 318

\noindent \textbf{Lovis C., Laskar J., et al., 2011,} A\&A, 528, A112

\noindent \textbf{Lynch, P. 2003,} MNRAS, 341, 1174

\noindent \textbf{McFadden L.A., Weissman P.R., Johnson T.V., 2007,} Encyclopedia of the Solar System. Academic Press, New York

\noindent \textbf{NASA Archive} https://exoplanetarchive.ipac.caltech.edu/

\noindent \textbf{Nieto M.M., 1972,} The Titius-Bode law of planetary distances: its history and theory. Pergamon Press, Oxford

\noindent \textbf{Nottale L., 1996,} Astron. Astrophys. 315, L09

\noindent \textbf{Nottale L., Schumacher G., Gay J., 1997,} Astron. Astrophys. 322, 1018

\noindent \textbf{Peale S.J., 1976,} Annual Rev.Astron.Astrophys., 14, 215

\noindent \textbf{PLATO} https://en.wikipedia.org/wiki/PLATO (spacecraft)

\noindent \textbf{Pletser V., 1988,} Earth Moon Planets 42, 118

\noindent \textbf{Pletser V., 2017,} Advances in Space Research, 60, 2314 [arXiv:1709.02704]

\noindent \textbf{Poveda A., Lara P., 2008,} Revista Mexicana de Astronomia y Astrofisica, 44, 243 [arXiv:0803.2240v1]

\noindent \textbf{Reinisch G., 1998,} Astron. Astrophys., 337, 299

\noindent \textbf{Richardson D.E., 1945,} Pop. Astron. 53, 14

\noindent \textbf{Saturn moons} https://en.wikipedia.org/wiki/\\ Moons\_of\_Saturn

\noindent \textbf{Scardigli F., 2007,} Found. Phys. 37, 1278 [arXiv:gr-qc/0507046]

\noindent \textbf{Schmidt O.J., 1945,} Comptes Rendus (Doklady) Acad. Sci. URSS 52, 667

\noindent \textbf{Scholkmann F., 2013,} Progress in Physics, 4, 85

\noindent \textbf{Scholkmann F., 2017,} Progress in Physics, 13 (2) 125

\noindent \textbf{Titius von Wittenberg J.D., 1766,} translation into German from French: Betrachtung über die Natur, vom Herrn Karl Bonnet (Johann Friedrich Junius, Leipzig), p. 7

\noindent \textbf{TOLIMAN} https://en.wikipedia.org/wiki/TOLIMAN

\noindent \textbf{Uranus moons} https://en.wikipedia.org/wiki/\\ Moons\_of\_Uranus

\noindent \textbf{Weizs\"acker C.F. von, 1943,} Zeit. f\"ur Astroph., 22, 319

\noindent \textbf{Willerding E., 1992,} Earth, Moon, and Planets, 56, 173
}

\appendix

\section{The statistical tools $R^2$ and $\chi^2$}
The statistical indicator $R^2$, called ``coefficient of determination", is defined in the following way. Given a set of points $(x_i,y_i)$, $i=1,...,N$, the ordinary least squares method tries to find the ``best" analytic function $y=f(x)$ that approximates the set $(x_i,y_i)$. Usually the function is a polynomial, and even more usually it is a straight line $y=A+Bx$. What one tries to minimize is the sum of the squares of the residuals
\be
\varrho = \sum_i [y_i - f(x_i)]^2 \,.
\label{sqerr}
\ee    
Of course, it would be useful to have ``something" with which to compare the sum $\varrho$, a sort of ``maximum error", in order to produce then a ``percentage error". The idea is then to consider the average of the $y_i$ data 
\be
\bar{y} = \frac{1}{N}\sum_i y_i
\ee
and to consider as maximum possible error the sum of the squares of the residuals in respect to this average $\bar{y}$
\be
\varrho_{max} = \sum_i (y_i - \bar{y})^2 \, .
\label{rhomax}
\ee
The percentage error is then clearly $\varrho/\varrho_{max}$, and the $R^2$ indicator is defined as
\be
R^2 = 1 - \frac{\varrho}{\varrho_{max}}\,.
\ee
Therefore, the agreement between the fitting line and the data is as better as $R^2$ is closer to 1.\\ 

The statistical indicator $\chi^2$ is defined, usually, as follow. Given a set of points $(x_i,y_i)$, $i = 1,...,N$, suppose that $y_i$ indicate the observed data (each $y_i$ is supposed to depend on the specific value $x_i$ of the parameter $x$); each $\hat{y}_i$ indicates the datum predicted by the (regression) model for the specific value $x_i$ of the parameter, namely $\hat{y}_i = f(x_i)$; each $\sigma_i$ indicates the uncertainty, or error bar, associated with the specific observed datum $y_i$. Then $\chi^2$ is defined as
\be
\chi^2 = \frac{1}{N}\sum_i\frac{(y_i - \hat{y}_i)^2}{\sigma_i^2}\,.
\ee 
In case of constant or uniform errors bars, namely $\sigma_i = \sigma$ for any $i$, then we can rescale $\sigma=1$, and 
\be
\chi^2 = \frac{1}{N}\sum_i(y_i - \hat{y}_i)^2 = \frac{\varrho}{N}\,,
\ee
and therefore 
\be
R^2 = 1 - \frac{N \chi^2}{\varrho_{max}}\,.
\ee
%

%

\section{The statistical tool $Median$}
The definition of \textit{Mean Absolute Deviation}, also known as \textit{Average Absolute Deviation} in Excel (where it is produced by the statistical function AVEDEV) is the following: given a set of numbers $\{x_1, x_2,\dots,x_N\}$ and defined the $Mean$ $\bar{x}$ in the usual way as 
\be
\bar{x} = \frac{1}{N}\sum_i x_i\,,
\ee 
then the \textit{Mean Absolute Deviation} referred to that set of numbers is
\be
MAD = \frac{1}{N} \sum_i |x_i - \bar{x}|
\ee
Clearly, MAD is a measure of how much ``close" the numbers $x_i$ are to their arithmetical mean $\bar{x}$. The closer to zero is MAD, the less ``dispersed" are the numbers $x_i$ around their mean $\bar{x}$. In the ideal case, when $MAD=0$ then $x_i=\bar{x}$ for any $i$.

In our context of (linear) regressions, we call $Median$ what is known as \textit{Mean Average Error} (MAE). Given a set of points $(x_i,y_i)$, $i=1,\dots,N$, where $y_i$ are the observed data (supposed to depend on the values $x_i$ of the parameter $x$), and $\hat{y}_i$ are the data predicted by the (regression) model, usually $\hat{y}_i=f(x_i)$, then the \textit{Mean Average Error} is defined as 
\be
MAE = \frac{1}{N}\sum_i|y_i - \hat{y}_i|\,.
\label{MAE}
\ee 
Obviously, the smaller is the $MAE$, the better is the fit by the model $y=f(x)$. The MAE can be compared with the \textit{Mean Squared Error}, MSE, another statistical indicator of the ``goodness" of a fitting model, defined as
\be
MSE = \frac{1}{N}\sum_i (y_i - \hat{y}_i)^2\,.
\ee
Of course, given the above definitions, the MSE and the sum of the squared residuals, Eq.\eqref{sqerr}, are linked by the relation
\be
\varrho = N \cdot MSE = \sum_i (y_i - \hat{y}_i)^2\,.
\ee 
%

\section{A theorem on the statistical indicators $R^2$, $Median$}

Given a planetary system, the relevant TB fitting can be constructed by using, as starting data, either the orbital periods $T_i$, namely the set of points $\{i,\log T_i\}$, or the semi-major axis $r_i$, namely the set of points $\{i,\log r_i\}$, with $i=1,2,3,\dots, N$. It is then quite easy to show that:
\be
R^2(T) = R^2(r)\,, \quad  {\rm Median}(T)=\frac{3}{2}\,{\rm Median}(r)  
\ee
\textit{namely, the $R^2$ of the data sets $\{i,\log T_i\}$ and $\{i,\log r_i\}$ are the same, while the Medians are proportional, through a factor $3/2$.}

\textbf{Proof}. By the Kepler Third Law we have $T_i^2 = k r_i^3$. Defining $Y_i=\log r_i$, $Z_i=\log T_i$, we can write
\be
Z_i \ = \ \frac{1}{2}\, \log k \ + \ \frac{3}{2}\, Y_i  \nonumber
\ee
and with reference to Eqs.\eqref{logTB}-\eqref{Z}, a straightforward calculation yields
\be
C \ = \ \frac{1}{2} \log k \ + \ \frac{3}{2}\, A\,; \quad \quad D \ = \ \frac{3}{2}\, B\,.  \nonumber
\ee 
\\
Defining now $\bar{Z}=(\sum_j Z_j)/N$ and $\bar{Y}=(\sum_j Y_j)/N$, we can then write
\be
\label{ZY}
Z_i - Z(i) &=& \frac{3}{2}(Y_i - Y(i))  \\
Z_i - \bar{Z} &=& \frac{3}{2}(Y_i - \bar{Y}) \nonumber
\ee 
and these two can be used together with Eqs.\eqref{sqerr},\eqref{rhomax} to arrive at
\be
\varrho(T) \ = \ \frac{9}{4}\, \varrho(r)\,, \quad \quad \varrho_{max}(T) \ = \ \frac{9}{4}\, \varrho_{max}(r)  \nonumber
\ee
which yield the thesis $R^2(T)=R^2(r)$.\\
Using again Eq.\eqref{ZY}, and the definition \eqref{MAE} of Median (i.e. MAE), we immediately also have
$Median(T)~=~(3/2)\,Median(r)$. QED.

\section{Titus-Bode rule for the satellites of Jupiter, Saturn, Uranus}

We show here how and to what extent the TB rule is obeyed by the systems of moons of, respectively, Jupiter, Saturn, and Uranus. Although this subject is obviously outside of the main topic of the present paper, from the historical point of view it presents a relevant interest, since it helps to depict the opinion that people formed about the TB law during the 20th century, before the discovery of exoplanetary systems. For the semi-major axis (or for the periods) of the satellites' orbits we use recent data taken from Wikipedia (see there the links Jupiter moons, Saturn moons, Uranus moons) and also from the \textit{Encyclopedia of the Solar System} (McFadden et al. (2007)).
In these tables and regression diagrams we do not include, on purpose, the satellites recently discovered only with spacecrafts (namely after Pioneer and Voyager missions), just because they are in general very light bodies, and it is well known that the TB rule works well only for quite large and quite massive objects (it does not work, for example, for comets or light 
asteroids). 
\begin{table}[h!]
\centering
\begin{tabular}{|c|c|c|}
	\hline
	Jupiter system & $\quad$ n $\quad$ & Orbital radius in km\\
	\hline
  Io       &  1  & 421,700\\
	Europa   &  2  & 671,034\\
	Ganymede &  3  & 1,070,412\\
	Callisto &  4  & 1,882,709\\
	\hline
\end{tabular}
\caption{Main satellites of Jupiter.}
\label{table11}
\end{table}

In the case of Jupiter, for example, the satellites Metis, Adrastea, Amalthea, and Thebe, all orbiting at distances shorter than the Io's orbit, are all bodies with sizes of order 100 km, and masses of $10^{-7} - 10^{-4}$ times the mass of Io. Therefore, for the above reasons, they are not considered in the regression diagram. \textit{A fortiori} the small moons orbiting at distances larger than Callisto orbit, all with sizes of 50 km or less, are ignored. The Jupiter diagram therefore displays only the Galileian satellites, Io, Europa, Ganymede, Callisto. 

Analogue considerations hold for the satellites systems of Saturn and Uranus. 

\begin{table}[h!]
\centering
\begin{tabular}{|c|c|c|}
	\hline
	Saturn system & $\quad$ n $\quad$ & Orbital radius in km\\
	\hline
  Mimas       &  1  & 185,539\\
	Enceladus   &  2  & 237,948\\
	Tethys      &  3  & 294,619\\
	Dione       &  4  & 377,396\\
	Rhea        &  5  & 527,108\\
	Titan       &  6  & 1,221,870\\
	Hyperion    &  7  & 1,481,009\\
	Iapetus     &  8  & 3,560,820\\
	\hline
\end{tabular}
\caption{Main satellites of Saturn.}
\label{table12}
\end{table} 
 
Hereafter the reader can find the regression diagrams for the main satellites of Jupiter, Saturn, and Uranus, each one equipped with the $R^2$ and $Median$ values computed from the specific statistical analysis. As the reader can easily realize, the statistical indicators $R^2$ and $Median$ assume really \textit{good} values, in particular for Jupiter and Uranus. In fact, both the Jupiter's and Uranus' $R^2$, as well as their $Median$s, are better than the Solar System $R^2$ and $Median$. Both Jupiter and Uranus systems fall in the set of $R^2>0.99$.

For the Saturn system the situation is clearly worse. This could perhaps be due to the fact that the system of Saturn contains more moons (146, May 2023) than any other planet in the Solar system. Moreover, the large rings might affect the distribution of the satellites' orbits in a still unknown way. 

In any case, it would be interesting to define a (precise) quantitative criterion to understand when a moon should be expected to obey a TB relation, or not. At present, the empirical selection rule (that we used to compile the tables \ref{table11},\ref{table12},\ref{table13}) is just based on the size/mass of the considered moon. We followed the tradition, since this rule is also adopted in the Nieto book, as well as in most of the literature on the Titius-Bode relation. Possible quantitative formulations of this criterion will form the subject of future works.       
\begin{table}[h!]
\centering
\begin{tabular}{|c|c|c|}
	\hline
	Uranus system & $\quad$ n $\quad$ & Orbital radius in km\\
	\hline
  Miranda       &  1  & 129,858\\
	Ariel         &  2  & 190,930\\
	Umbriel       &  3  & 265,982\\
	Titania       &  4  & 436,282\\
	Oberon        &  5  & 583,449\\
	\hline
\end{tabular}
\caption{Main satellites of Uranus.}
\label{table13}
\end{table}  
\newpage
\begin{figure}[h]
\includegraphics[scale=0.6]{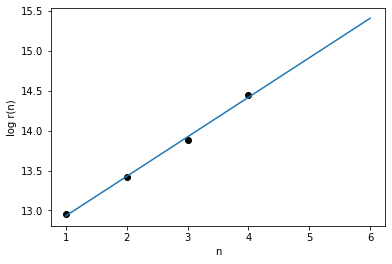}
\caption{Fitting of the main satellites of Jupiter. The statistical indicators are $R^2=0.99759458$ and $Median = 0.00181383$.}
\label{Jupiter_moons.png}
\end{figure}
\begin{figure}[h!]
\includegraphics[scale=0.6]{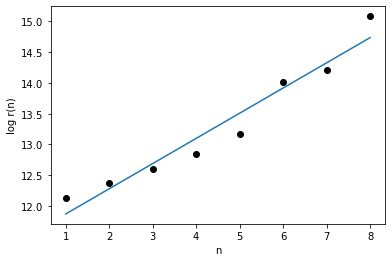}
\caption{Fitting of the main satellites of Saturn. The statistical indicators are $R^2=0.94507382$ and $Median = 0.014263996$.}
\label{Saturn_moons.png}
\end{figure}
\begin{figure}[h!]
\includegraphics[scale=0.6]{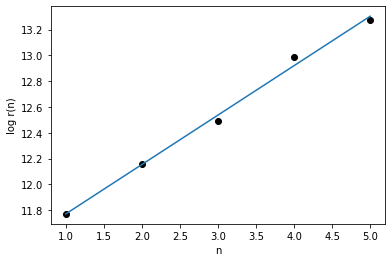}
\caption{Fitting of the main satellites of Uranus. The statistical indicators are $R^2=0.99512428$ and $Median = 0.00204344$.}
\label{Uranus_moons.png}
\end{figure}
\newpage
%
\section{References for the planetary systems in Table \ref{table1}}

\begin{table}[!ht]
\centering
\begin{tabular}{|c|c|l|}
	\hline
	System & Planets & References \\
	\hline
    Sun &  9 + Ceres  & https://doi.org/10.1016/C2010-0-67309-3 \\
	\hline
	KOI-351 & b,c,d,e,f,g,h & https://doi.org/10.1088/0004-637X/781/1/18 \\
	        & i             & https://doi.org/10.3847/1538-3881/aa9e09 \\
	\hline				
	Trappist 1 & b,c,d,e,f,g,h & https://doi.org/10.3847/PSJ/abd022 \\
	\hline
	HD 10180 &  b,c,d,e,f,g  &  https://doi.org/10.1088/0004-637X/792/2/111 \\
	\hline
	HD 191939 &  b,c,d,e,f,g  &  	https://doi.org/10.1051/0004-6361/202244120 \\
	\hline
	HD 219134 &  b,c,d,e  &  https://doi.org/10.1038/s41550-017-0056 \\
	          &  f,g      &  https://doi.org/10.1088/0004-637X/814/1/12 \\
	\hline
	HD 34445 & b  &  https://doi.org/10.3847/1538-3881/aa5df3 \\
	         & c,d,e,f,g & https://doi.org/10.3847/1538-3881/aa8b61 \\
	\hline
	K2-138 &  b,c,d,e,f,g  &  	https://doi.org/10.1051/0004-6361/201936267 \\
	\hline
	Kepler 11 & b,c,d,e,f,g & https://doi.org/10.1088/0004-637X/770/2/131 \\
	\hline
	Kepler 80 &  b,c,d,e,f & https://doi.org/10.3847/0004-6256/152/4/105 \\
	          &   g        & https://doi.org/10.3847/1538-3881/aa9e09 \\
	\hline
	TOI - 1136 & b,c,d,e,f,g & https://doi.org/10.3847/1538-3881/ad1330 \\
	\hline
	TOI - 178 & b,c,d,e,f,g & https://doi.org/10.1051/0004-6361/202039767 \\
	\hline
	HD 108236 & b,c,d,e,f & https://doi.org/10.1051/0004-6361/202039608 \\
	\hline
	HD 158259 &  b,c,d,e,f  & 	https://doi.org/10.1051/0004-6361/201937254 \\
	\hline
	HD 23472 &  b,c,d,e,f  & 	https://doi.org/10.1051/0004-6361/202244293 \\
	\hline
	HD 40307 & b,c,d,e,f & 	https://doi.org/10.1051/0004-6361/201220268 \\
	\hline
	K2-268 & b,c,d,e,f &  https://doi.org/10.1093/mnras/stab2305 \\
	\hline
	K2-384 & b,c,d,e,f & https://doi.org/10.3847/1538-3881/ac5c4c \\
	\hline
	Kepler 102 & b,c,d,e,f  &  	https://doi.org/10.1051/0004-6361/202346211 \\
	\hline
	Kepler 122 & b,c,d,e &  https://doi.org/10.1088/0004-637X/784/1/45 \\
	           &  f      &  https://doi.org/10.1088/0004-637X/787/1/80 \\
	\hline
	Kepler-150 & b,c,d,e  & https://doi.org/10.1088/0004-637X/784/1/45 \\
	           &  f       & https://doi.org/10.3847/1538-3881/aa62ad \\
	\hline
	Kepler 154 & b,c & https://doi.org/10.1088/0004-637X/784/1/45 \\
	           & d,e,f &  https://doi.org/10.3847/0004-637X/822/2/86 \\
	\hline
	Kepler 169 & b,c,d,e,f &  https://doi.org/10.1088/0004-637X/784/1/45 \\
	\hline
	Kepler 186 & b,c,d,e &  https://doi.org/10.1126/science.1249403 \\
	           &  f      &  https://doi.org/10.1088/0004-637X/800/2/99 \\
	\hline
	Kepler 238 & b,c,d & https://doi.org/10.1088/0004-637X/784/1/45 \\
	           &  e,f  & https://doi.org/10.1088/0067-0049/210/2/25 \\
	\hline
	Kepler 292 & b,c,d,e,f &  https://doi.org/10.1088/0004-637X/784/1/45 \\
	\hline
	Kepler 32 & b &  https://doi.org/10.3847/1538-3881/abd93f \\
	          & c,d,e,f  &  https://doi.org/10.1088/0004-637X/750/2/114 \\
	\hline
	Kepler 33 & b,c,d,e,f &  https://doi.org/10.1088/0004-637X/750/2/112 \\
	\hline
  Kepler 55 & b,c & https://doi.org/10.1093/mnras/sts090 \\
	          & d,e,f  &  https://doi.org/10.1088/0004-637X/784/1/45 \\
	\hline
Kepler 62 & b,c,d,e,f &  https://doi.org/10.1126/science.1234702 \\
	\hline
Kepler 82 & b,c,f &  https://doi.org/10.1051/0004-6361/201935879 \\
          & d,e   &  https://doi.org/10.1088/0004-637X/784/1/45  \\
	\hline
Kepler 84 & b,c &  https://doi.org/10.1088/0067-0049/208/2/22 \\
          & d,e,f &  https://doi.org/10.1088/0004-637X/784/1/45  \\
	\hline
\end{tabular}
\caption{References for planetary systems in Table \ref{table1}.}
\label{table14}
\end{table}
\end{document}